\documentclass{aa}

\usepackage{graphicx}
\usepackage{txfonts,longtable,dcolumn,verbatim}    

\usepackage[colorlinks=true,urlcolor=blue,citecolor=red,linkcolor=red,bookmarks=true]{hyperref}

\usepackage{amsmath,amssymb,url}
\usepackage{gensymb}

\begin{document}

\title{Debiased population of very young asteroid families} 

\author{D. Vokrouhlick\'y\inst{1},
        D. Nesvorn\'y\inst{2},
        M. Bro{\v z}\inst{1},
        W.F. Bottke\inst{2}}

\titlerunning{Debiased population of very young asteroid families}
\authorrunning{Vokrouhlick\'y et~al.}

\institute{Institute of Astronomy, Charles University, V~Hole\v{s}ovi\v{c}k\'ach 2,
           CZ-180~00 Prague 8, Czech Republic \\ \email{vokrouhl@cesnet.cz}
      \and
           Department of Space Studies, Southwest Research Institute,
           1050 Walnut St., Suite 300, Boulder, CO 80302, USA}

\date{Received: \today ; accepted: October 20, 2023}

\abstract
 {Asteroid families that are less than one million years old offer a unique possibility to
 investigate recent asteroid disruption events and test ideas about their dynamical evolution.
 Observations provided by powerful all-sky surveys have led to an enormous increase  in the number
 of detected asteroids over the past decade. When the known populations are well characterized,
 they can be used to determine asteroid detection probabilities, including those in young families,
 as a function of their absolute magnitude.}
 {We use observations from the Catalina Sky Survey (CSS) to determine the bias-corrected population of
 small members in four young families down to sizes equivalent to several hundred meters.}
 {Using the most recent catalog of known asteroids, we identified members from four young
 families for which the population has grown appreciably over recent times. A large fraction of
 these bodies have also been detected by CSS. We used synthetic populations of asteroids, with their
 magnitude distribution controlled by a small number of parameters, as a template for the bias-corrected model
 of these families. Applying the known detection probability of the CSS observations, we could
 adjust these model parameters to match the observed (biased) populations in the young families.}
 {In the case of three families, Datura, Adelaide, and Rampo, we find evidence that the magnitude
 distribution transitions from steep to shallow slopes near  $300$ to
 $400$~meters. Conversely, the Hobson family population may be represented by a single power-law 
 model. The Lucascavin family has a limited population; no new members have been
 discovered over the past two decades. We consider a model of parent body rotational fission 
 with the escaping secondary tidally split into two components (thereby providing three
 members within this family). In support of this idea, we find that no other asteroid with absolute
 magnitude $H\leq 18.3$ accompanies the known three members in the Lucascavin family. A similar
 result is found for the archetypal asteroid pair Rheinland--Kurpfalz.}
{}


\maketitle


\section{Introduction} \label{intr}
More than a century ago, \citet{hira1918} discovered the first examples of
statistically significant clusters in the space of asteroid heliocentric orbital
elements (using proper values of the semimajor axis, eccentricity, and inclination).
Suspecting their mutual relation, he coined the term asteroid families. Hirayama 
rightly proposed that the families are collections of asteroids related to parent bodies
that disrupted sometime in the past. He even identified asteroid collisions as
the source  of these catastrophic events. Over time, asteroid families became a core
element of Solar System small body science. They provide (i) an important constraint
on asteroid collisional models; (ii) a unique tool to study the internal structure of
large asteroids, both in terms of their chemical homogeneity and mechanical structure;
(iii) an important source of impactor showers that include both large projectiles and
dust onto the surfaces of the terrestrial planets (including the Earth); (iv) an arena
for studying a plethora of dynamical processes affecting the orbits and spins of asteroids;
and (v) many more \citep[see, e.g., recent reviews by][]{netal2015,maetal2015,
mietal2015,bojanrev2022}. 

In this paper, we explore (ii), namely the capability of asteroid family data to constrain
the internal structure of the parent body. Over the past two decades or so, sophisticated
numerical approaches have been developed to model energetic asteroidal collisions,
the subsequent dispersal, and gravitational re-accumulation of resulting fragments
\citep[e.g.,][]{mietal2015,aetal2015,juetal2015}.
The outcome of these simulations, which may be compared to the information provided by 
asteroid families, sensitively depends on assumptions about the internal properties of the
parent body. One type of dataset includes the size frequency distribution of asteroid members
in the family. While determining asteroid family members looks straightforward, it has potential
complications. This is because many families extend over non-negligible portions of the
asteroid belt. As a result, the proper zone in orbital element space in which the family
members are located may contain a certain fraction of unrelated (interloping) asteroids. Methods
to estimate the interloper fraction have been developed \citep[e.g.,][]{mi1995}, but their
validity is limited and their results are necessarily of a statistical, rather than deterministic,
nature. Additionally, progress from powerful and automated surveys over the past decades makes it
more difficult to deal with the interloper problem because small asteroid spatial densities
increasely fill proper element space. Unless we know the size distribution of the background and
the family population, more asteroids mean that there are more interlopers to deal with.  

Fortunately, the ability of surveys to increase the known asteroid populations has also brought
into play a new and interesting niche that allows us to determine the complete (bias-corrected)
population of the family members. The fundamental goal of this paper is to try to exploit this
possibility. Our focus here is on a special subclass of asteroid families characterized by
extremely young ages, namely those that are $\simeq 1$~Myr or less. Already the first examples,
which were discovered little less than two decades ago \citep{daturaSci2006,nv2006}, help us 
understand the
means to get rid of potential interlopers. Consider that the unusual youth of these families
means that five of the six osculating orbital elements are clustered (semimajor axis $a$,
eccentricity $e$, inclination $I$, and longitudes of node $\Omega$ and perihelion $\varpi$), rather
than the standard three proper orbital elements used for most family work (semimajor axis $a_{\rm p}$,
eccentricity $e_{\rm p}$, and inclination $I_{\rm p}$). This immediately has two positive consequences.
First, our work can use simpler osculating elements rather than less (population-wise) accessible
proper elements. Second, the additional two dimensions of the orbital element arena in which we
searched for these very young families have a diluted spatial density of known asteroids. The very
young families show up as distinct, and often isolated clusters, allowing us to largely circumvent
the interloper problem. Additionally, their recent origin has allowed us to accurately determine each
family’s age by propagating the asteroid orbits backward in time and then by observing
how the orbits rearrange themselves into a tighter cluster at the epoch of its formation. 
This procedure has helped to further eliminate interlopers. 

As far as the population count is concerned, we are then left with the observational bias produced
by telescopic limitations (basically the capability of a given instrument to detect asteroids to
some apparent magnitude). Here we can compensate for this problem to a degree by using asteroids taken
from a well-characterized and sufficiently long-lasting survey. Profiting from our earlier work,
in which we developed a new model for the near-Earth asteroid population, we use a careful
characterization of the Catalina Sky Survey ($1.5$-m Mt.~Lemmon telescope, G96) observations
in between 2016 and 2022. We apply this rich dataset to determine the bias-corrected population of 
four very young families  and a few more clusters of interest.%
\footnote{The first attempt of the method has been carried out by \citet{daturaAA2017}, who applied it to
 the case of Datura family. However, both (i) the precise detection efficiency characterization of the
 older set of Catalina survey observations, and (ii) mainly the Datura family known population, were
 significantly smaller than in the present paper.}

We first briefly describe the observation set in Sec.~\ref{obs}. Next, we introduce the very young
families that we are going to analyze in this paper (Sec.~\ref{yf}), providing their new identification
and full membership in the Appendix. In Sec.~\ref{res} we develop an approach to determine the complete
population of the families, based on their biased population and information about the survey detection
probability, and we apply it to the selected cases. In Sec.~\ref{concl} we discuss the implications of
our results and provide some discussion of  potential future work.

\section{Catalina Sky Survey observations} \label{obs}
Catalina Sky Survey%
\footnote{\url{https://catalina.lpl.arizona.edu/}}
(CSS), managed by Steward Observatory of the University of Arizona,
has been one of the most prolific survey programs over the past decade
\citep[e.g.,][]{CSSEPSC2019}. While primarily
dedicated to the discovery and further tracking of near-Earth objects with
the goal to characterize a significant fraction of the population with sizes as small
as $140$~m, CSS observations represent an invaluable source of information for
other studies in planetary science or astronomy in general.
\begin{table*}[ht]
\caption{\label{orient}	
 Asteroid families and pairs studied in this paper.}
\centering
\begin{tabular}{clccl}
\hline \hline
   & Name & $N_{\rm obs}$ & $N_{\rm CSS}$ & Goal \\ 
\hline	
             \multicolumn{5}{c}{\it \rule{0pt}{3ex} -- Very young families 1 --} \\                    
\rule{0pt}{3ex}
 AF & Datura   & 91 & 60 & Parameters of the bias-corrected population \\
 AF & Adelaide & 79 & 63 & Parameters of the bias-corrected population \\
 AF & Hobson   & 60 & 33 & Parameters of the bias-corrected population \\
 AF & Rampo    & 42 & 26 & Parameters of the bias-corrected population \\ [6pt]
             \multicolumn{5}{c}{\it -- Very young families 2 --} \\                    
\rule{0pt}{3ex}
 AF & Wasserburg  & 8 & 8 & Possibly steep distribution of small fragments \\
 AF & Martes      & 6 & 3 & Possibly steep distribution of small fragments \\  [6pt]
             \multicolumn{5}{c}{\it -- Special: starving families and pairs --} \\                    
\rule{0pt}{3ex}
 AF & Lucascavin           &  3 & 3 & No additional members? \\
 AP & Rheinland/Kurpfalz   &  2 & 2 & No additional members? \\  [2pt] 
\hline
\end{tabular}
\tablefoot{Very young families 1 with abundant
 population of known members allow us to estimate parameters of the bias-corrected population.
 Very young families 2 contain smaller numbers of known members; here we can only indicate
 steep progression of currently unobservable members. Starving families and asteroid pairs
 contain up to three members only; here we aim at disproving additional members with
 absolute magnitude smaller than some threshold.
 The first column identifies the asteroid category: AF for the asteroid family, AP for
 the asteroid pair. The second column provides the name, the third and fourth column give number
 of known members and number of members detected by Catalina Sky Survey. The last column states
 in brief our goals in this work.}
\end{table*}

Here, we use observations of the CSS 1.5-m survey telescope located at Mt. Lemmon (MPC observatory
code G96). Our method builds on the work of \citet{nes2023}, who constructed
a new model of the near-Earth object population using CSS data. They carried out a detailed analysis
of the asteroid detection probabilities for the G96 operations over the period between January 2013
and June 2022. This interval was divided into two phases: (i) observations before May~14, 2016
(phase~1), and (ii) observations after May~31, 2016 (phase~2).  The first phase contained $61,585$
well-characterized frames, in the form of sequences of four that were typically $30$~s exposure
images, while the second phase had $162,280$ well-characterized frames. The reason for the difference
was due to longer timespan of the phase 2 but also
an important upgrade of the CSS CCD camera in the second half of May 2016. The new camera had four
times the field of view, and better photometric sensitivity, allowing the survey 
to cover a much larger latitude region about the ecliptic. The superiority of the CSS observations
taken during phase~2 allows us to drop the phase~1 data in most of the work below. Only in the
case of Lucascavin family and Rheinland-Kurpfalz pair do we combine
observations from the two phases into a final result.

The final product of interest for our work here is the detection probability $p(H)$ as a
function of the absolute magnitude $H$ for asteroids in a chosen family. In principle, $p$ 
depends not only on $H$, but on all orbital elements (in other words, it is specific to
a particular body). Members in the youngest known asteroid families to date, however, have
their orbit longitudes $\lambda$ uniformly distributed in between $0^\circ$ and $360^\circ$.
This is because the characteristic $\lambda$ dispersal timescale after the family forming event
is only about $1$-$3$~kyr; all families which we consider here are at least an order
of magnitude older than this value. Conversely, a property of 
very young asteroid families are that they have a tight clustering in the other five orbital
elements, including the longitude of node $\Omega$ and longitude of perihelion $\varpi$.
As a result, the detection probability $p(H)$ assigned to a given family has been computed
using the mean values of osculating orbital elements, except for $\lambda$ where
the individual probabilities have been averaged.%
\footnote{We used 10,000 synthetic orbits characteristic to the family and $\lambda$ uniformly
 distributed in its definition interval to determine the mean value of $p(H)$.}
Only in the case of two families --Datura and Rampo-- we used the secular angles $\Omega$ and
$\varpi$ to randomly sample their observed interval of values shown in Figs.~\ref{fd_1}
and \ref{fr_1}. As seen in those figures, and expected from theoretical considerations, the 
$\Omega$ vs. $\varpi$ values are strongly correlated in very young families. We take this
correlation into account when computing the mean detection probability $p(H)$. 

Apart from 
$p(H)$, we can also determine a detection rate $r(H)$, namely a statistically mean number
of the survey fields of view in which a given family member with an absolute magnitude $H$
should have been detected. While correlated with $p(H)$, $r(H)$ contains additional information
and may be thus used as a consistency check in our analysis below. Technical details of the
numerical methods that allow us to determine $p(H)$ and $r(H)$ can be found in
\citet{nes2023}.

\section{Very young families} \label{yf}
In this section we introduce four very young asteroid families, namely Datura, Adelaide, Hobson
and Rampo, whose known population is large enough that they are  
suitable candidates for our debiasing efforts.%
\footnote{Obviously, a second criterion of their selection is that CSS detected a significant
 fraction of known members in these families during its phase~2 operations.}
We also consider two additional families, Wasserburg and Martes, that have extremely young ages
but whose population is limited. For these examples, we do not perform a full-scale debiasing
analysis but instead argue that a large population
of small undetected members should exist near the currently known population.
Finally, we consider two special cases: the very young asteroid family Lucascavin and
the asteroid pair Rheinland--Kurpfalz. Here, our goal is actually opposite to the previous
cases. Our working hypothesis is that further smaller fragments in their location might not exist.
As a result, we use CSS observations to set an upper limit on the size or magnitude of the unseen members to
explore whether this  hypothesis might be correct.
\begin{figure}[t]
 \begin{center} 
  \includegraphics[width=0.47\textwidth]{f1.eps}
 \end{center}
 \caption{Osculating values of the secular angles -- longitude of node $\Omega$ 
  (abscissa) and longitude of perihelion $\varpi$ (ordinate) -- for members in the
  Datura family (epoch MJD 60,000.0). The black symbols show multi-opposition orbits,
  gray symbols are for the singleopposition orbits; diamond symbol for (1270) Datura,
  the largest member. Larger/smaller relative values of the secular angles, measured
  with respect to (1270) Datura, are correlated with positive/negative shift in proper
  semimajor axis $\Delta a_{\rm P}$ (as explained by the arrows). Because the node drifts
  in a retrograde sense, while the perihelion drifts in the prograde sense, their
  trends are opposite to each other (thence anticorrelation of the two angles). Location of
  the exterior mean motion resonance M9/16 with Mars is mapped where the label shows.
  The dashed line with a slope $-0.5$ indicates the (anti-) correlation trend. The symbol
  indicated by a question mark shows projection of (429988) 2013~PZ36 (an object captured
  on a very chaotic possibly in the exterior mean motion resonance E3/10 with Earth), whose
  association to the Datura family is uncertain.}
 \label{fd_1}
 \end{figure}
\begin{figure}[t]
 \begin{center} 
  \includegraphics[width=0.47\textwidth]{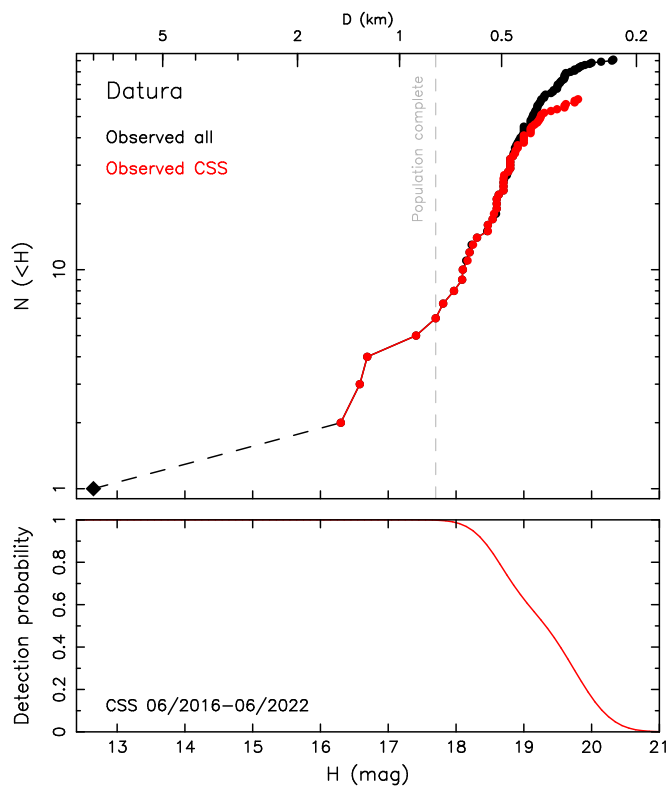}
 \end{center}
 \caption{Top panel: Cumulative magnitude distribution $N(<H)$ of the Datura family members.
  The black symbols are all $91$ known members (including the largest asteroid (1270) Datura
  shown by the diamond, but disregarding (429988) 2013~PZ36, whose membership in the family is
  uncertain); the red symbols are $60$ members detected by CSS during the phase~2
  operations. The top abscissa indicates an approximate size computed from $H$ with an
  assumption of $p_V=0.24$ value of the geometric albedo. Bottom panel: Detection probability
  $p(H)$ of Datura members as a function of $H$ during the phase~2 operations of CSS based on
  analysis of geometric and photometric detection factors run on a large synthetic population
  of Datura members. We find that $p=1$ up to $H\simeq 18$ magnitude, which sets the limit
  where the Datura population is complete (dashed line on the upper panel). Beyond this
  limit $p$ decreases to zero at about $21$ magnitude.}
 \label{fd_2}
 \end{figure}

Table~\ref{orient} provides a basic overview of the asteroid clusters and pairs that are analyzed
in this paper, as well as some notes on the goals we hope to achieve. The identification method
used to find the families, and full listing of the family members for each family analyzed in this
paper, is provided in the Appendix. In what follows, we provide basic information about the
investigated families, with slightly more attention paid to the Datura family. 
The debiasing procedure to constrain a complete population of members in the
above-mentioned families is presented in the next Sec.~\ref{res}.

\subsection{Very young families with large population of members}\label{largevyf}

\noindent{\it Datura.-- }The cluster of asteroids about the largest member (1270) Datura
is an archetype of very young families. In this sense it is comparable to the Karin family,
which is an excellent example of a sizable young family having an age less than $\simeq 10$~Myr
but secular angles distributed uniformly in the $0^\circ$ to $360^\circ$ interval. Not only is
the Datura family the first example discovered in the very young family class \citep[][see
also \citet{nv2006}]{daturaSci2006}, but its location in the inner part of the main belt
allowed us to readily collect the physical parameters of the largest members and study the
role of the very young families in a broader context of planetary science 
\citep[e.g.,][]{mdn2008,ver2009,vetal2008,daturaAA2009,daturaAA2017}.
The number of known members in the Datura family has also grown quickly from only 7 in
2006 to 17 in 2017. 

Here we make use of the accelerating pace with which asteroids have
been discovered during recent years and report a currently known Datura population of
$N_{\rm obs}=91$ members (possibly even 94 members, see Table~\ref{datura_members}).
Importantly, $N_{\rm CSS}=60$ of them has been also detected by CSS during its phase~2
operations. We note that \citet{daturaAA2017} already attempted to use CSS observations
for their Datura population debiasing efforts. Our current work, however, surpasses the
detail and accuracy of this earlier work. \citet{daturaAA2017} could use only the 13
largest members in the Datura family detected by CSS between 2005 and 2012. Thanks to
the camera update by CSS in 2016, the six years of CSS operations between 2016 and
2022 has led to a much larger Datura population and an improved characterization of
family member detection probabilities.
\begin{figure}[t]
 \begin{center} 
  \includegraphics[width=0.47\textwidth]{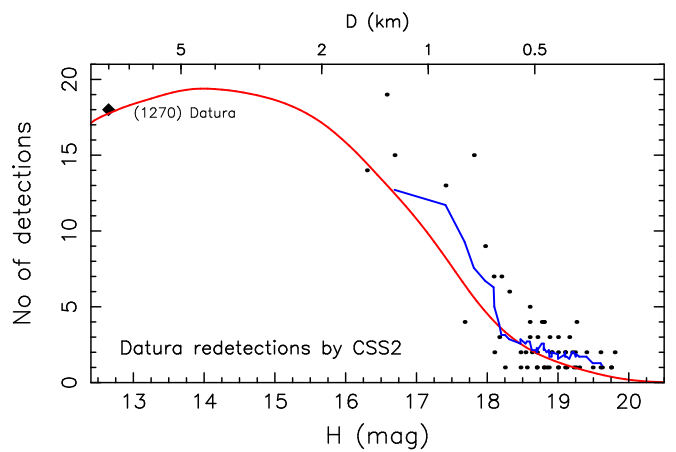}
 \end{center}
 \caption{Number of detections of the identified Datura family members during the phase 2
  CSS operations: the largest body (1270) Datura shown by a diamond symbol and highlighted
  using a label, other $59$ smaller members shown by black symbols. The red line is the
  theoretical prediction based on a large synthetic Datura population computed together with
  the detection probability $p(H)$ from Fig.~\ref{fd_2}. The blue curve is a mean number
  of detections for the observed Datura members computed on a running window of 7 consecutive
  data-points.}
 \label{fd_3}
 \end{figure}
\begin{figure}[t]
 \begin{center} 
  \includegraphics[width=0.47\textwidth]{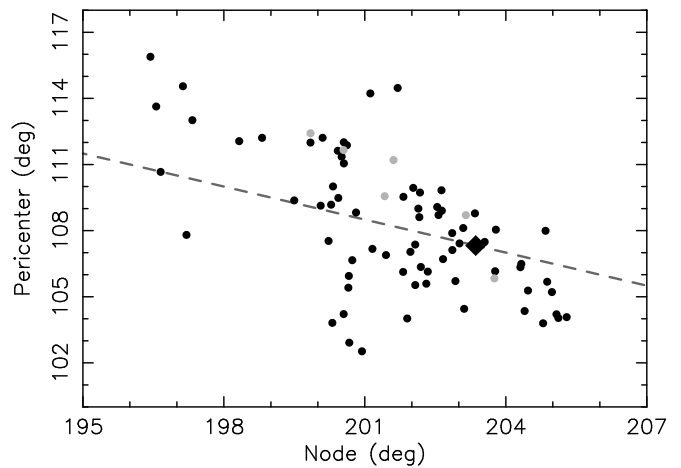}
 \end{center}
 \caption{Osculating values of the secular angles -- longitude of node $\Omega$ 
  (abscissa) and longitude of perihelion $\varpi$ (ordinate) -- for members in the
  Adelaide family (epoch MJD 60,000.0). The black symbols show multiopposition orbits,
  gray symbols are for the singleopposition orbits; diamond symbol for (525) Adelaide,
  the largest member. The dashed line has the expected slope $-0.5$ (see, e.g., Figs.~\ref{fd_1}
  and \ref{fr_1} for Datura and Rampo families), but the data are more scattered in the
  Adelaide case. This is likely due to Mars perturbation discussed in \citet{adelaideAA2021}.}
 \label{fa_1}
 \end{figure}
\begin{figure}[t]
 \begin{center} 
 \includegraphics[width=0.47\textwidth]{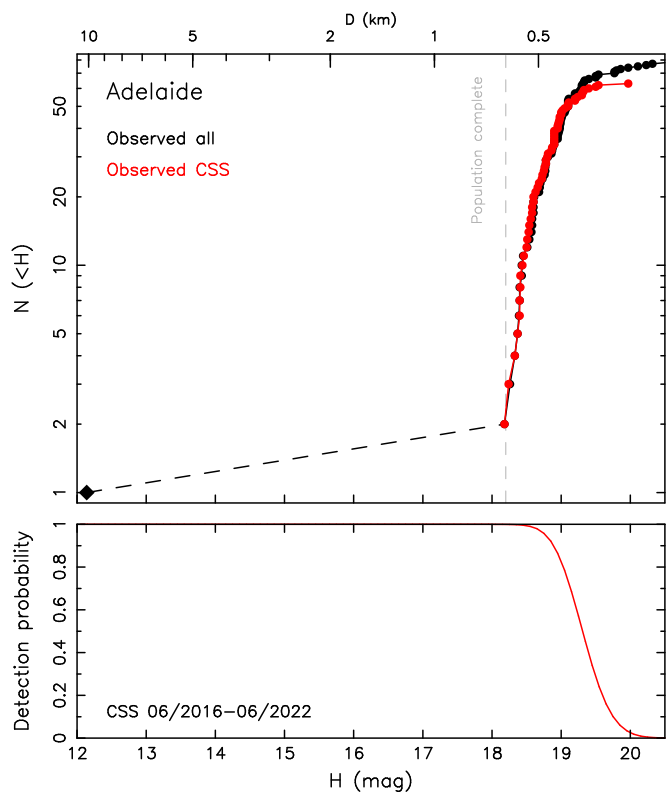}
 \end{center}
 \caption{Top panel: Cumulative magnitude distribution $N(<H)$ of the Adelaide family members.
  The black symbols are all $79$ known members (including the largest asteroid (535) Adelaide
  shown by the diamond), the red symbols are $63$ members detected by CSS during the phase~2
  operations. The top abscissa indicates an approximate size computed from $H$ with an
  assumption of $p_V=0.24$ value of the geometric albedo. Bottom panel: Detection probability
  $p(H)$ of Adelaide members as a function of $H$ during the phase~2 operations of CSS based on
  analysis of geometric and photometric detection factors run on a large synthetic population
  of Adelaide members. We find that $p=1$ up to $H\simeq 18.2$ magnitude, which sets the limit
  where the Adelaide population is complete (dashed line on the upper panel). Beyond this
  limit $p$ decreases to zero at about $20.4$ magnitude.}
 \label{fa_2}
 \end{figure}
\begin{figure}[t]
 \begin{center} 
 \includegraphics[width=0.47\textwidth]{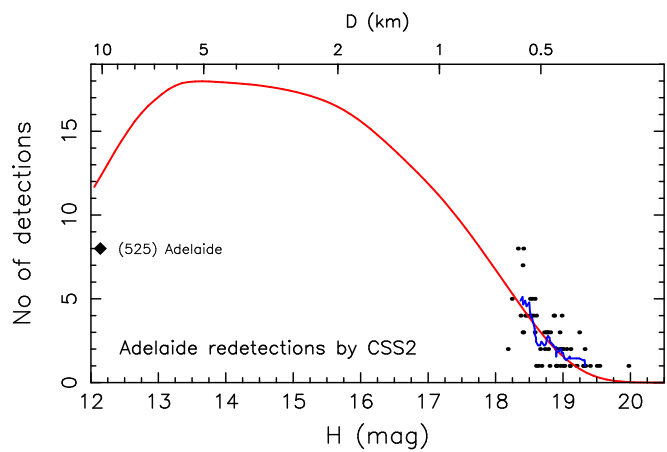}
 \end{center}
 \caption{Number of detections of the identified Adelaide family members during the phase 2
  CSS operations: the largest body (525) Adelaide shown by a diamond symbol and highlighted
  using a label, other $62$ smaller members shown by black symbols. The red line is the
  theoretical prediction based on a large synthetic Adelaide population computed together with
  the detection probability $p(H)$ from Fig.~\ref{fa_2}. The blue curve is a mean number
  of detections for the observed Adelaide members computed on a running window of 9 consecutive
  data-points.}
 \label{fa_3}
 \end{figure}

Before we turn our attention to the magnitude distribution of the Datura members, we
use this family to exemplify some common features of very young clusters. They help to
justify membership of given asteroids within the family, even without a further
substantiation via a detailed study of their past orbital convergence using numerical
integrations (see a brief discussion of this issue in the Appendix).
A correlation between the osculating values of the secular angles, namely longitude of node
$\Omega$ and longitude of perihelion $\varpi$, is a characteristic property of several
very young families (unless the family is extremely young, such that $\Omega$ and
$\varpi$ are clustered within a degree or so, basically corresponding to their initial dispersal).
Denoting $\Delta \Omega$ and $\Delta \varpi$ as the angular difference with respect to the
largest body in the family, the initial phase of the dispersal process is described
by a linear approximation. Thus at time $T$, one has $\Delta \Omega(T)\simeq C \,T +
{\cal O}(T^2)$ and a similar equation for the  longitude of perihelion, with $C=(\partial s/
\partial a)\,\Delta a$, where $s$ is the proper nodal frequency and $\Delta a$ is the
difference in semimajor axis with respect to the largest body produced by the initial
velocity ejection. The smallest observed fragments in Datura have $\Delta a\simeq 2\times
10^{-3}$ au, corresponding to their ejection by $\simeq 10$ m~s$^{-1}$ (only slightly 
larger than the escape velocity from the parent body of the family). Together with
$(\partial s/\partial a)\simeq 40$ arcsec~yr$^{-1}$~au$^{-1}$, we can estimate their
angular difference $\Delta \Omega\simeq 11^\circ$ in $T\simeq 500$~kyr (see Fig.~\ref{fd_1}).
A similar analysis for $\Delta \varpi$ results in about half this value. 

Given that in
both $\Delta \Omega$ and $\Delta \varpi$ the nonlinear terms in time $T$ are still
very small \citep[those will be produced by the Yarkovsky drift in semimajor axis of
the small members in the family; e.g.,][]{daturaAA2009,daturaAA2017}, they are strongly
correlated with a slope $-0.5$. The early dispersal phase of very young families
is characterized by additional correlations between the osculating elements, namely (i)
the eccentricity and longitude of perihelion, and (ii) the inclination and the longitude
of node (see, e.g., data in the Tables given in the Appendix). As mentioned above,
these extra correlations between osculating orbital elements help to strengthen
justification of the membership in the family.

The cumulative magnitude distribution $N(<H)$ of Datura family members is shown in 
Fig.~\ref{fd_2}. The magnitudes $H$ for the six lowest-numbered members were determined
accurately using calibrated observations, and expressed at the mid-value of the lightcurve,
by \citet{daturaAA2009} and \citet{daturaAA2017}. The magnitudes for other Datura members
were taken from the MPC catalog. We show both the distribution of all known members (black
symbols), and highlight also the sample of $60$ members which have been detected by CSS
(red symbols). Data for these asteroids may be used for debiasing of the Datura family
population, since only for them we have the detection probability well characterized. 

The bottom panel 
of Fig.~\ref{fd_2} shows detection probability $p(H)$ of Datura members as a function of their
absolute magnitude. As explained above, this is a result based on an analysis of 10,000 synthetic
orbits in the  Datura family zone  which makes $p(H)$ very smooth. At the first sight, it 
might be surprising that $p\simeq 1$ up to magnitude $18$, signaling that the population of the 
family members is complete up to that limit. This inference, however, is correct and a result of
(i) a six year  survey , (ii) the small value of Datura-like orbital inclinations, such that
CSS fields-of-view did not miss an opportunity to detect the asteroids in the Datura family,
and (iii) a typical 50\% photometric detection limit of CSS in between 20.5 and 21.5 
apparent magnitude (in the visual bands). Neglecting a small phase-angle correction in the Pogson's
relation between absolute $H$ and apparent $m$ magnitudes, we have $H\simeq m-5 \log(r\,\Delta)$,
where $r$ and $\Delta$ are heliocentric and geocentric distances of the asteroid. At opposition,
and near aphelion to cover the worst case situation, we have $r\simeq 2.7$~au and $\Delta\simeq
1.7$~au. As a result, the limiting magnitude $m\simeq 21$ translates to $H\simeq 17.8$.
During the 6~yr period of CSS phase~2, the configuration eventually becomes favorable
to detection, explaining the completion limit at $H\simeq 18$ magnitude. At the opposite end of things, the
probability $p$ has a tail to nearly $21$ magnitude. This means CSS with its best nightly limits
near the apparent $22$ magnitude have a chance to detect small Datura members when they happen 
to be near the perihelion of their orbit at opposition.
\begin{figure}[t]
 \begin{center} 
 \includegraphics[width=0.47\textwidth]{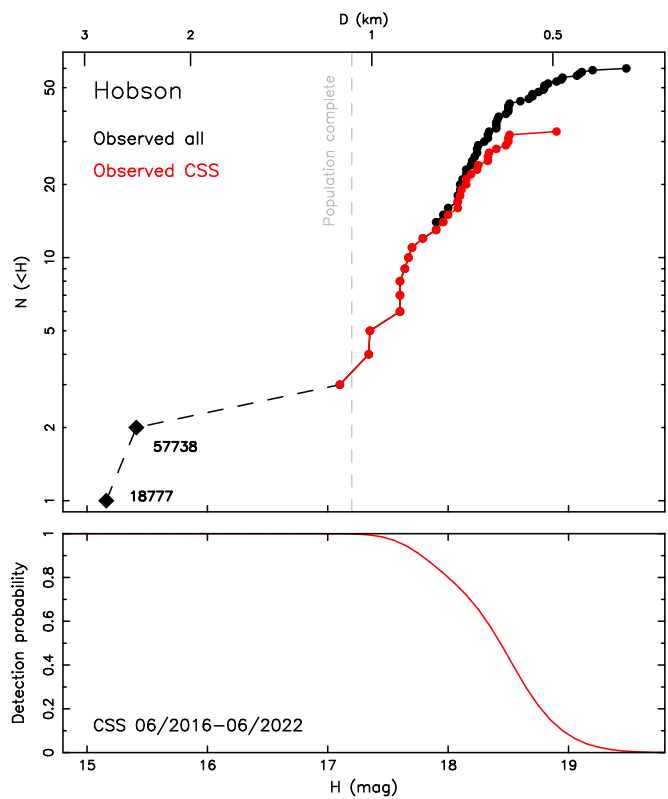}
 \end{center}
 \caption{Top panel: Cumulative magnitude distribution $N(<H)$ of the Hobson family members.
  The black symbols are all $60$ known members (including the largest asteroids (18777) Hobson
  and (57738) 2001~UZ160 shown by the diamond), the red symbols are $33$ members detected by
  CSS during the phase~2
  operations. The top abscissa indicates an approximate size computed from $H$ with an
  assumption of $p_V=0.2$ value of the geometric albedo. Bottom panel: Detection probability
  $p(H)$ of Hobson members as a function of $H$ during the phase~2 operations of CSS based on
  analysis of geometric and photometric detection factors run on a large synthetic population
  of Hobson members. We find that $p=1$ up to $H\simeq 17.2$ magnitude, which sets the limit
  where the Hobson population is complete (dashed line on the upper panel). Beyond this
  limit $p$ decreases to zero at about $19.5$ magnitude.}
  \label{fh_1}
 \end{figure}

In order to verify that the detection probability $p(H)$ shown in Fig.~\ref{fd_2} is
reasonable, we also determined the expected mean rate $r(H)$ of CSS phase~2 detections 
and compared it with the actual number of detections of all $60$ identified Datura members.
This result is presented in Fig.~\ref{fd_3}. 
The largest body (1270) Datura was found to be  detected 18 times,
and even members up to magnitude $H=18$ were typically detected more than 10 times. This
is a good verification of  population completeness. Only after that limit does the number of detections
decrease,  with no Datura member having $H>20$ magnitude detected. This outcome 
corresponds to the inferred detection probability: $p<0.1$ for $H>20$. 

The mean value
of the actual Datura-member detections computed using a running window of consecutive 7 asteroids
is shown by the blue curve. The scatter of the number of detections about the predicted red line
is not surprising because the latter has been computed as a mean value from 10,000 synthetic
Datura members. The important point is that the blue curve, though computed as a mean over a
much smaller number of cases (additionally having different $H$ values), reasonably follows the
predicted mean rate. This points to consistency in evaluation of the detection probability too.
\smallskip

\noindent{\it Adelaide.-- }The cluster of five small objects about the inner main belt asteroid
(525) Adelaide was first reported by \citet{ade2019}. Apart from an approximate age of $500$~kyr,
few  details were given in this paper. \citet{car2020}, while trying to search for
secondary subclusters in the very young asteroid families, analyzed the Adelaide family and identified
19 of its members. The case was finally revisited by \citet{adelaideAA2021}, who noted a
significant population increase to about $50$ small asteroids in this family. They confirmed the earlier
age estimate and considered a possibility of a causal link between formation of the Datura and 
Adelaide families (which they rejected). \citet{bojanrev2022} identified already $72$ members, and
our current census of the Adelaide family population reveals $N_{\rm obs}=79$ members, yet another
important increase. The population increase rate of the Adelaide family is among the largest of the
very young families. A fortunate circumstance for our analysis is that $N_{\rm CSS}=63$ of the
members were also detected by CSS in its phase~2.

Figure~\ref{fa_1} shows the osculating secular angles $\Omega$ and $\varpi$ of the Adelaide
family population from Table~\ref{adelaide_members} in the Appendix. The $\Omega$ vs. $\varpi$ 
correlation is weaker than that of the Datura-family members (Fig.~\ref{fd_1}).
\citet{adelaideAA2021}, while analyzing behavior of the backward propagated orbits in the
Adelaide family, noted a weak chaotic signature triggered by a conjoint effect of weak mean-motion
resonances and distant encounters with Mars. We suspect they are also the origin of the
observed scatter in the correlation between the secular angles seen in Fig.~\ref{fa_1}.
Nevertheless, the orbits show a high degree of clustering even in the subspace of the
secular angles, which in effect strengthens their membership in the family.

The cumulative magnitude distribution of the Adelaide members is shown in Fig.~\ref{fa_2}.
Its extreme behavior has been already noted by \citet{adelaideAA2021}: (i) the largest
remnant (525) Adelaide is separated from other members in the family by an unusually large
gap of 6 magnitudes in the absolute magnitude $H$ scale, and (ii) the small fragment population
has an extremely steep $H$-distribution in between $18$ and $19$ (the local power-law
$N(<H)\propto 10^{\gamma H}$ approximation requires $\gamma\simeq 2$ or even larger). This shape
is characteristic of a cratering event on (525) Adelaide. 

The bottom panel in
Fig.~\ref{fa_2} shows the mean detection probability $p(H)$ of the Adelaide members. 
The range in which $p(H)$ drops from one to zero, namely $H\simeq 18.4$ to $H\simeq 20.4$
magnitudes, is narrower than in the case of the Datura family (with the completion limit
at even higher magnitude). This is readily explained by a smaller eccentricity of
Adelaide-like orbits for nearly the same value of semimajor axis and inclination.
\begin{figure}[t]
 \begin{center} 
 \includegraphics[width=0.47\textwidth]{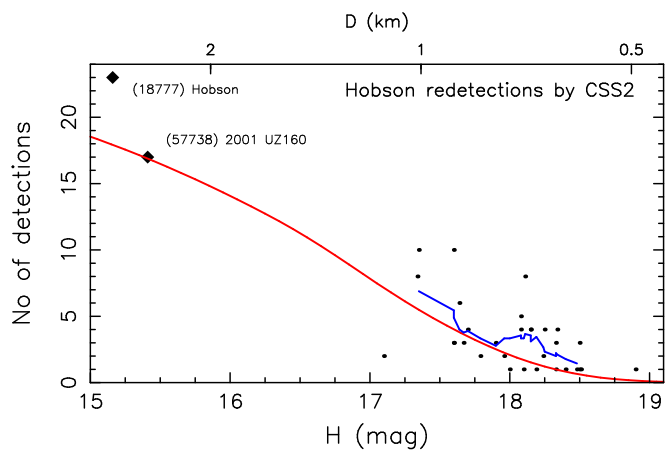}
 \end{center}
 \caption{Number of detections of the identified Hobson family members during the phase 2
  CSS operations: the largest bodies (18777) Hobson and (57738) 2001~UZ160 shown by a diamond
  symbol and highlighted
  using a label, other $31$ smaller members shown by black symbols. The red line is the
  theoretical prediction based on a large synthetic Hobson population computed together with
  the detection probability $p(H)$ from Fig.~\ref{fh_1}. The blue curve is a mean number
  of detections for the observed Hobson members computed on a running window of 9 consecutive
  data-points.}
  \label{fh_2}
 \end{figure}

To further check our results, we also compared the number of CSS phase~2 detections of the
$63$ Adelaide members and their mean computed rate $r(H)$ (Fig.~\ref{fa_3}). The largest 
asteroid (525) Adelaide has been detected $8$ times, which conforms --within fluctuation– to
the predicted rate of about $13$. We note the decrease of $r(H)$ for objects brighter than
magnitude $13$. This phenomenon in the CSS observations has to do with the
saturation of the signal for bright objects, as they can become confused with stationary
sources hiding their sky-plane motion. Such a configuration may occasionally happen when
(525) Adelaide is at opposition near perihelion of its orbit. Small members then sample
the tail of $r(H)$ values with only few detections predicted. The running mean of 
detections (blue curve) appears to follow the predicted $r(H)$ dependence reasonably well.
\smallskip

\noindent{\it Hobson.-- }\citet{pv2009} identified a small cluster of asteroids associated with
the largest member (57738) 2001~UZ160 and set an upper age of $500$~kyr for its formation event.
They also noted a nearby asteroid (18777) Hobson, but were unsure about its relation to the
cluster, mainly because Hobson and 2001~UZ160 have similar sizes, which they considered unusual
for the outcome of a collisional fragmentation of the parent body. \citet{rp2016,rp2017,rp2018}
then revisited the situation and proved that Hobson was associated with the cluster. 
They derived an age for the family of $365\pm 67$~kyr. 
By 2018, their Hobson population consisted of nine members, which shortly 
improved to 11 by the work of \citet{petal2018}. These latter authors also rejected the possibility
of the Hobson family formation by rotation fission, and conducted valuable photometric
observations of the two largest members Hobson and 2001~UZ160. The two
similar-size largest remnants also intrigued \citet{hobsonAA2021}, who revisited the nature of the
parent object of this family (counting already 45 Hobson members, and \citet{bojanrev2022}
reported another increase to 51 members). Using the SPH/N-body formation simulation, their
results implied a very special impact and target combination was required. As a novel idea,
they also argued the Hobson family may result from collisional fragmentation of a component
in a parent binary. In this work we report $N_{\rm obs}=60$ members (likely even one more, see
Table~\ref{hobson_members}),
out of which $N_{\rm CSS}=33$ were detected during the phase~2 of CSS operations.
\begin{figure}[t]
 \begin{center} 
 \includegraphics[width=0.47\textwidth]{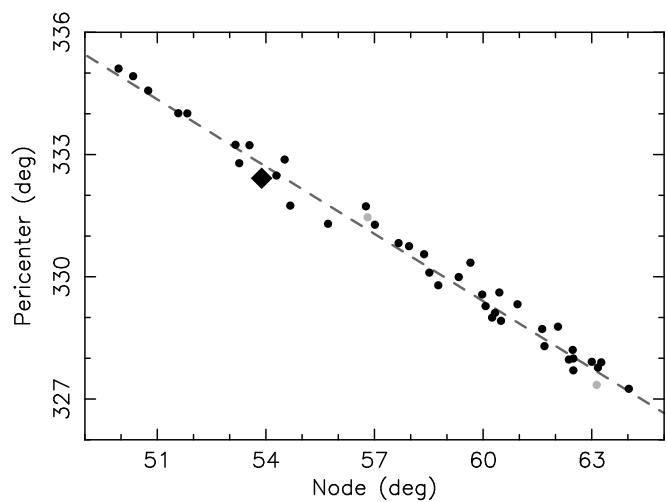}
 \end{center}
 \caption{Osculating values of the secular angles -- longitude of node $\Omega$ 
  (abscissa) and longitude of perihelion $\varpi$ (ordinate) -- for members in the
  Rampo family (epoch MJD 60,000.0). The black symbols show multiopposition orbits,
  gray symbols are for the singleopposition orbits; diamond symbol for (10321) Rampo,
  the largest member. Larger/smaller relative values of the secular angles, measured
  with respect to (10321) Rampo, correlated with positive/negative shift in proper
  semimajor axis. The node/perihelion
  trends are opposite, because the node drift in a retrograde sense, while the
  perihelion drifts in the prograde sense. The dashed line with a slope
  $-0.5$ indicates the correlation trend.}
 \label{fr_1}
 \end{figure}

The dispersion of the secular angles within about two degrees is
a consequence of the very young age of the Hobson family. We thus turn our attention
directly to the cumulative magnitude distribution of its members shown on Fig.~\ref{fh_1}.
The two largest asteroids --(18777) Hobson and (57738) 2001~UZ160-- are its most outstanding
feature. Their orbital convergence has been independently verified by \citet{rp2017,rp2018}
and \citet{hobsonAA2021}, while \citet{petal2018} determined the identical values of the 
$V-R$ color index (compliant with the S-type taxonomy). As a result, 
their membership to the cluster appears to be solid. 

The bottom panel on Fig.~\ref{fh_1} shows the mean detection probability $p(H)$ determined
for the CSS phase~2 operations. It appears similar to that of the Datura family except for
about a magnitude shift towards smaller $H$, which implies completion down to $H\simeq 17.1$
magnitude. This result is easily understood by a comparison with the Datura family; Hobson's
family has similar eccentricity and inclination values, but a larger set of semimajor axis
values. The Hobson family resides in the central part of the main asteroid
belt next to the J3/1 mean motion with Jupiter.
\begin{figure}[t]
 \begin{center} 
 \includegraphics[width=0.47\textwidth]{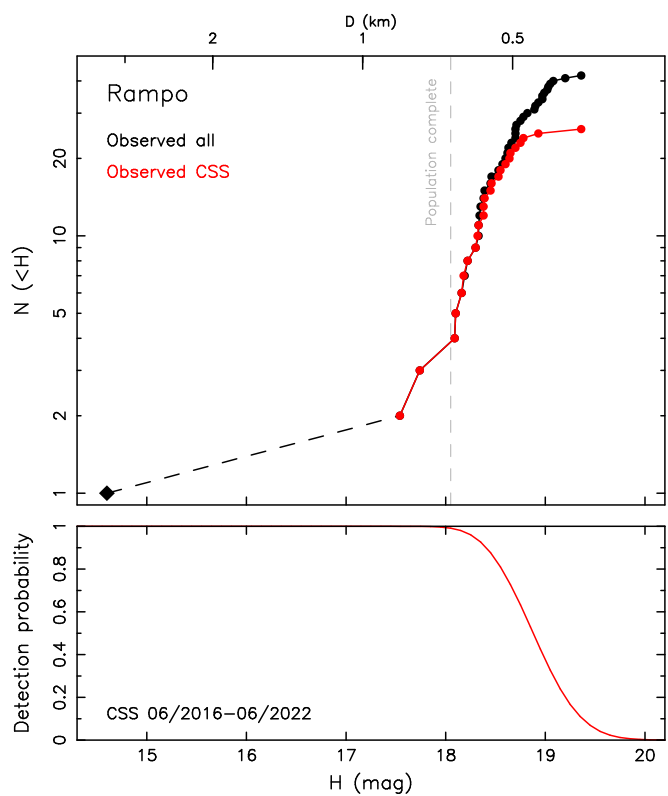}
 \end{center}
 \caption{Top panel: Cumulative magnitude distribution $N(<H)$ of the Rampo family members.
  The black symbols are all $42$ known members (including the largest asteroid (10321) Rampo
  shown by the diamond), the red symbols are $26$ members detected by CSS during the phase~2
  operations. The top abscissa indicates an approximate size computed from $H$ with an
  assumption of $p_V=0.24$ value of the geometric albedo. Bottom panel: Detection probability
  $p(H)$ of Rampo members as a function of $H$ during the phase~2 operations of CSS based on
  analysis of geometric and photometric detection factors run on a large synthetic population
  of Rampo members. We find that $p=1$ up to $H\simeq 18$ magnitude, which sets the limit
  where the Rampo population is complete (dashed line on the upper panel). Beyond this
  limit $p$ decreases to zero at about $20$ magnitude.}
  \label{fr_2}
 \end{figure}

The inferred mean rate of detections $r(H)$ for the phase~2 of CSS matches, within the
statistical fluctuations, the actual number of detections of Hobson members
(Fig.~\ref{fh_2}). The brightest two asteroids stand out with more than 15 detections, while 
members in the small-size tail typically have fewer than five detections.
\smallskip

\noindent{\it Rampo.-- }The core of this family, namely two small asteroids tightly clustered
about (10321) Rampo, has been found by \citet{pv2009}. Focusing on asteroid pairs, these
authors reported a probable age between $0.5$ and $1.1$~Myr. About a decade later,
\citet{petal2018} discovered another four small members in this family and used
backward orbital integration to assess a more accurate age of $780^{+130}_{-90}$~kyr.
Finally, \citet{bojanrev2022} revisited the Rampo family population and identified
$36$ small members around the largest remnant (10321) Rampo. Here we find the Rampo
family population has increased to $N_{\rm obs}=42$ (possibly even $44$, see 
Table~\ref{rampo_members} in the Appendix); $N_{\rm CSS}=26$ of them were
detected during CSS phase~2.

The correlation of the secular angles $\Omega$ and $\varpi$, shown in Fig.~\ref{fr_1},
is exemplary among the very young families. The family must be still in the dispersion regime
that is linear with time (i.e., the same discussed for the Datura family). Similarly to
the Datura case, the orbits of Rampo family members exhibit strong correlations in the 
pairs of orbital elements $e$ vs. $\varpi$ and $I$ vs. $\Omega$, providing us with a useful 
justification for their family membership.

\begin{figure}[t]
 \begin{center} 
  \includegraphics[width=0.47\textwidth]{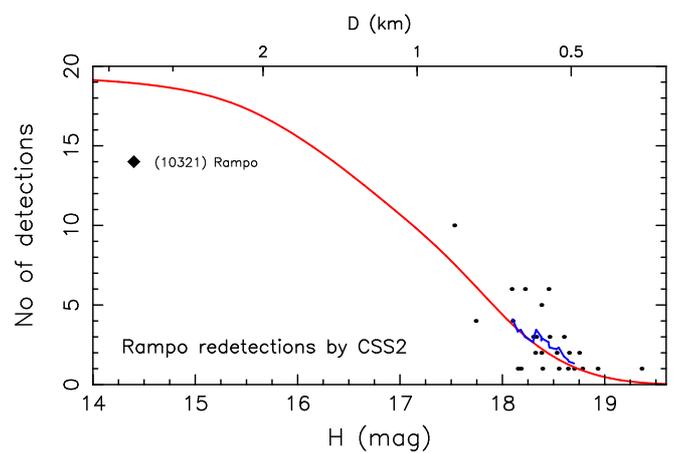}
 \end{center}
 \caption{Number of detections of the identified Rampo family members during the phase 2
  CSS operations: the largest body (10321) Rampo shown by a diamond symbol and highlighted
  using a label, other $25$ smaller members shown by black symbols. The red line is the
  theoretical prediction based on a large synthetic Rampo population computed together with
  the detection probability $p(H)$ from Fig.~\ref{fr_2}. The blue curve is a mean number
  of detections for the observed Rampo members computed on a running window of 9 consecutive
  data-points.}
  \label{fr_3}
 \end{figure}

The cumulative magnitude distribution of Rampo family members shares some similarities with
the Datura cluster; compare Figs.~\ref{fr_2} and \ref{fd_2}. The small differences
consist of: (i) a larger magnitude gap $\Delta H$ between the largest members and the
second largest member ($\Delta H\simeq 3.8$ for Datura and $\Delta H\simeq 3.2$ for 
Rampo), and (ii) a larger size of (1270) Datura over (10321) Rampo
\citep[by about a factor $\simeq 2.15$ accounting for a slight albedo difference][both
being S-class spectral taxonomy]{petal2018}. Similar to Datura, the former feature suggests 
that the family may have been formed by a large cratering event, though more work on
this issue is required \citep[e.g.,][]{d2007}. The Rampo members have a detection
probability $p(H)$
computed for phase~2 of CSS transitions that go from one at $H\simeq 18$ to zero
at $H\simeq 20$. This sharp transition is due to their small eccentricities. The completion
limit is similar for both families because their aphelion distances are comparable
(on the other hand, the perihelion distance is smaller for Datura orbits and thus
its $p(H)$ reaches to larger absolute magnitudes). As in all cases discussed in this
paper, the number of CSS phase~2 detections of Rampo
family members nicely follows the predicted rate $r(H)$ (Fig.~\ref{fr_3}).

\subsection{Extremely young asteroid families with small number of known members}\label{smallvyf}

\noindent{\it Wasserburg.-- }A very tight asteroid pair of two Hungaria objects (4765) Wasserburg
and 2001~XO105 was reported by  \citet{vn2008}. \citet{pra2010}, analyzing the formation
process of asteroid pairs, included this couple in their sample and reported an approximate
age larger than $90$~kyr. \citet{petal2019}, compiling the most detailed
study of the asteroid pair population, noted a small asteroid 2016~GL253 accompanying the pair
on a very close orbit and suggested the trio of asteroids may be the large-end tip  of
a very young family in the Hungaria population. \citet{bojanrev2022} confirmed the trend, detecting
six members in what they called the Wasserburg family. Here we find two more members
in the family, completing the count at $N_{\rm obs}=8$. Interestingly, all of them were also
detected during phase~2 of the CSS operations, thence $N_{\rm CSS}=8$.
\begin{figure}[t]
 \begin{center} 
 \includegraphics[width=0.47\textwidth]{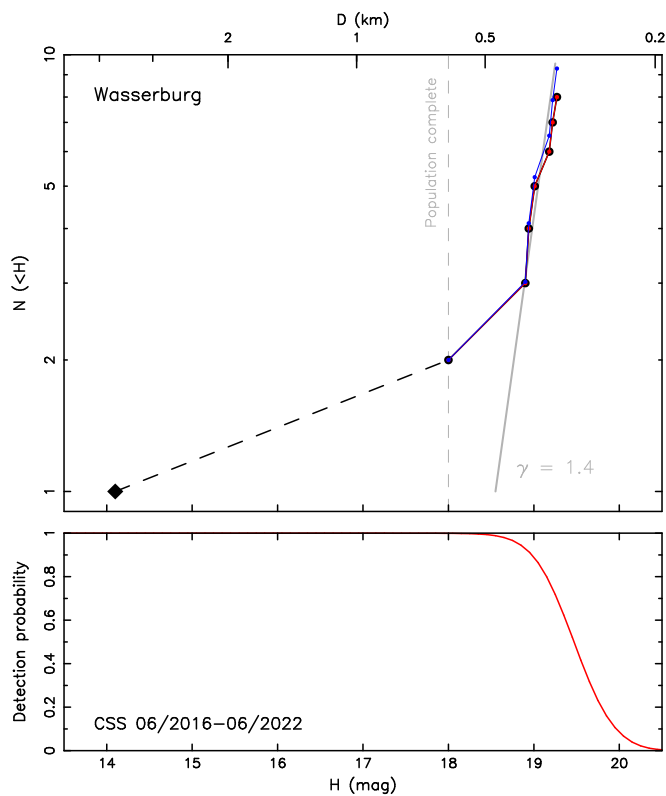}
 \end{center}
 \caption{Top panel: Cumulative magnitude distribution $N(<H)$ of the Wasserburg family members.
  The black symbols are all $8$ known members (including the largest asteroid (4765) Wasserburg
  shown by the diamond), the blue symbols provide the simplest variant of the debiased 
  population. This was obtained by incrementing the population by $1/p(H_{i+1})$, when stepping from
  absolute magnitude $H_{i}$ to $H_{i+1}$ ($i=1,\ldots, N_{\rm obs}$); the observed (biased)
  population increments by one by definition. The gray line is an approximate local power-law representation
  $N(<H)\propto 10^{\gamma\,H}$ near $H\simeq 19$ with $\gamma\simeq 1.4$. The top abscissa indicates an
  approximate size computed from $H$ with an
  assumption of $p_V=0.3$ value of the geometric albedo. Bottom panel: Detection probability
  $p(H)$ of Wasserburg members as a function of $H$ during the phase~2 operations of CSS based on
  analysis of geometric and photometric detection factors run on a large synthetic population
  of Wasserburg members. We find that $p=1$ up to $H\simeq 18.3$ magnitude, which sets the limit
  where the Wasserburg population is complete (dashed line on the upper panel). Beyond this
  limit $p$ decreases to zero at about $20.5$ magnitude. }
 \label{fw_1}
 \end{figure}

The cumulative magnitude distribution of the presently known members of the Wasserburg
family is shown in Fig.~\ref{fw_1}. The bottom panel on the same figure provides the
detection probability $p(H)$ during CSS phase~2 operations. The completion
limit is near $H\simeq 18.5$ magnitude, impressively large in spite of the high
inclination of the Wasserburg family orbits (being part of the Hungaria zone).
Some of these orbits may be missed by the fields-of-view of CSS. The situation improved
after July~2016, however, with the wide field camera reaching
well beyond the $\pm 30^\circ$ zone around the ecliptic. So the geometric losses
are small, and the heliocentric proximity of the Hungaria region helped to detect
even small asteroids. Indeed, the six smallest members in the Wasserburg family
have an  absolute magnitude near or even above the $H=19$ limit. 

As mentioned in the preamble
of this Section, the small number of identified members in this family
does not permit a full-scale debiasing effort. Accordingly, we only conducted the
simplest estimate to characterize the complete Wasserburg population using the
following steps:
\begin{itemize}
\item We considered the observed (biased) population of the family members and
 sorted their absolute magnitude values $\{H_i\}$, with $i=1,\ldots, N_{\rm obs}$,
 from the smallest to the largest value; 
\item By definition, the observed population increases by one when shifting
 along the list according to the ordered $H$-values; we assume the largest member
 in the family is bright enough such that $p(H_1)=1$;
\item The simplest estimate of the complete population is then obtained by again
 moving along the vector $\{H_i\}$ of ordered absolute magnitudes, but now
 incrementing the population by $1/p(H_i)$ instead of one.
\end{itemize}
The result is shown by the blue curve at the top panel of Fig.~\ref{fw_1}. Since
even the smallest Wasserburg fragment has $p(H_8)\simeq 0.71$ (in other words,
detection of even the smallest known fragments is 
expected), the complete population does not deviate too much from the observed
population. Up to that point the cumulative magnitude distribution is very steep,
locally approximated by a power law with an exponent of $\gamma\simeq 1.4$.
This value is only slightly shallower than that observed in the case of the Adelaide
family. From that similarity, we may tentatively conclude that Wasserburg family
has resulted from a huge cratering event in (4765) Wasserburg itself, though again 
there are many additional possibilities \citep[e.g.,][]{d2007}.
\begin{figure}[t]
 \begin{center} 
 \includegraphics[width=0.47\textwidth]{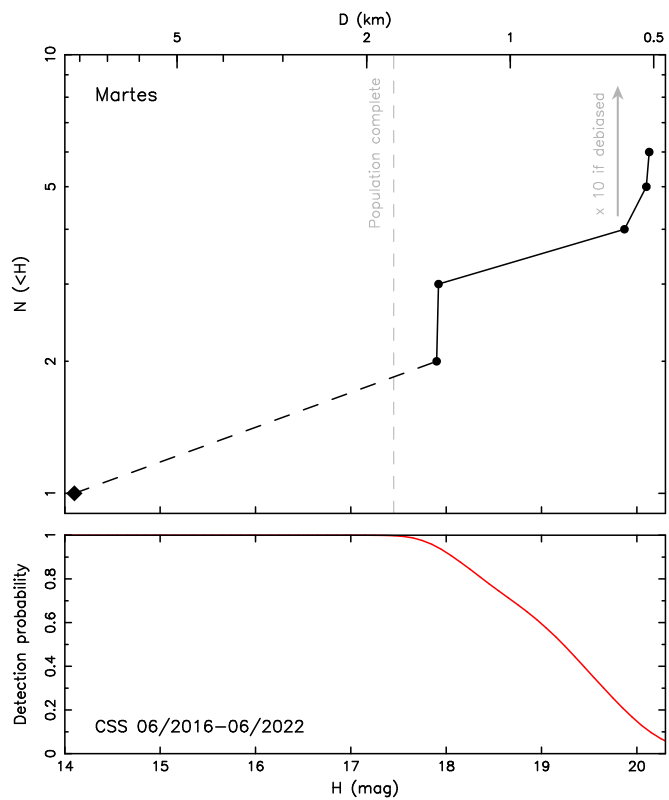}
 \end{center}
 \caption{Top panel: Cumulative magnitude distribution $N(<H)$ of the Martes family members.
  The black symbols are all $6$ known members (including the largest asteroid (5026) Martes
  shown by the diamond). The top abscissa indicates an approximate size computed from $H$ with an
  assumption of $p_V=0.06$ value of the geometric albedo (conforming the Ch-class
  taxonomy). Bottom panel: Detection probability
  $p(H)$ of Martes members as a function of $H$ during the phase~2 operations of CSS based on
  analysis of geometric and photometric detection factors run on a large synthetic population
  of Martes members. We find that $p=1$ up to $H\simeq 17.5$ magnitude, which sets the limit
  where the Martes population is complete (dashed line on the upper panel). Beyond this
  limit $p$ decreases to zero at about $20.5$ magnitude. At magnitude $\simeq 20$
  $p\simeq 0.1$. This implies that the three very small members recently detected must
  represent a tiny sample of a much larger population having about the same size.}
  \label{fm_1}
 \end{figure}

However, an outstanding puzzle here is to explain why the current surveys have yet to 
detect any smaller fragments. This reason is because of the inferred steepness
of the magnitude distribution, and the non-negligible detection probability
$p(H_8)$ mentioned above. In other words, a fair number of the subsequent members
in the Wasserburg family should have a detection probability $\simeq 0.5$, yet 
none have been detected. Does this mean that the magnitude distribution beyond the
detected population suddenly becomes shallow. The answer to this question is left for 
future analysis.
\smallskip
\begin{figure}[t]
 \begin{center} 
 \includegraphics[width=0.47\textwidth]{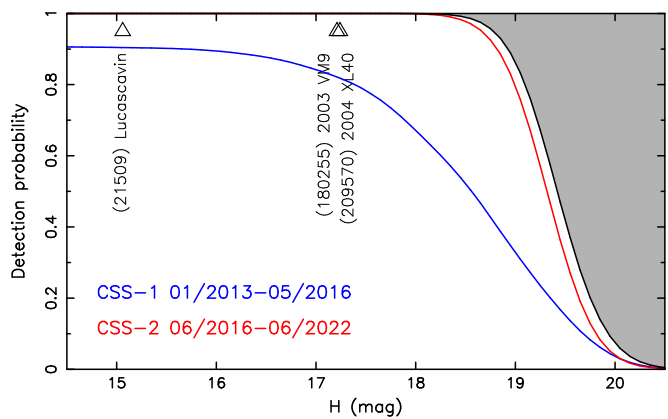}
 \end{center}
 \caption{Detection probability of an additional small fragment in the Lucascavin
  family: (i) the blue curve is $p_1$ during the phase~1 of CSS operations, and
  (ii) the red curve is $p_2$ during the phase~2 of CSS operations. The black curve is the
  combined probability $p$ during both phases (Eq.~\ref{comb12}). The gray area allows
  the existence of an additional small body in the system, whose maximum probability of
  occurrence is complementary value to the probability $p$ on the left ordinate.}
 \label{ls_1}
 \end{figure}

\noindent{\it Martes.-- }\citet{vn2008} mentioned (5026) Martes and 2005~WW113 among their
list of tight asteroid pairs. As also noted by these authors, some of these pairs were
expected to be the two largest members in a collisionally born asteroid family (e.g., 
Wasserburg family). Recently, \citet{bojanrev2022} reported a third member in the tight orbital
region about Martes, namely 2010~TB155, while our census in this paper increases the number by
three more small objects, with  $N_{\rm obs}=6$ (Table~\ref{martes_members}), with the last
three asteroids associated with the Martes cluster discovered in Autumn 2022.%
\footnote{All three of them were pre-covered on CCD images taken by Pan-STARRS in
 2011, and also detected in 2014 by a 4-m Victor M. Blanco telescope on Cerro Tololo,
 using the Dark Energy Camera, which can reach much fainter objects than the 1.5-m
 G96 telescope.}
Only the largest three members in the Martes family were detected by CSS, such that
$N_{\rm CSS}=3$. The Martes cluster is a part of a much larger Erigone family, whose age
has been estimated to $\simeq 280$~Myr \citep[e.g.,][]{yy2006,spot2015} or $130\pm 30$~Myr
by \citet{bot2015}. This association is justified by the objects having the same spectral
taxonomic type Ch as the Erigone family and (5026) Martes \citep{poli2014}. The extremely
clustered orbital elements of the Martes members suggest an unusually young age for the 
family. Indeed, \citet{petal2019} found $18\pm 1$~kyr, a slight improvement on the result 
of \citet{pra2010}. We find that the orbits of the smallest
three members may also converge to this time window, further justifying the Martes age,
but a detailed analysis would need to consider the thermal accelerations in the
simulation. We leave this effort to a separate study, but conclude here that the
Martes family has the youngest currently known age.

Figure~\ref{fm_1} shows the absolute magnitude distribution of the Martes family
members. Admittedly this distribution is an incomplete portion of the family population, and
for that reason we do not attempt a serious debiasing effort.
We only note the behavior of the detection probability $p(H)$ determined for the phase~2
operations of CSS (bottom panel on Fig.~\ref{fm_1}). Martes-family orbits have the
largest eccentricity among our sample, and this produces the largest stretch of $H$ 
values in which $p(H)$ decreases from 1 to 0. At magnitude $H\simeq 20$ we have
$p\simeq 0.1$. 

Taken at a face value, we would infer a large population of small members in the Martes
family, such that every one of the three may represent in fact $\simeq 1/p\simeq 10$ 
asteroids. This  logic might be  flawed, however, because the three small members
were not  detected by phase~2 CSS. Strictly speaking, 
we should not use them to infer anything about Martes family magnitude distribution.
Nevertheless, we believe our inferences may be close to reality. This is because all
three smallest asteroids in the Martes family were  detected by G96/CSS in
September 2022. This time period is technically out of the phase~2 interval, but only by a
small amount. It also shows the capability of G96 to detect them. The size of the Martes
population at $H\simeq 20$ is left for future work.

\subsection{Starving young asteroid families with only three known members and asteroid pairs}
\label{starv}

\noindent{\it Lucascavin.-- }This very tight cluster of three asteroids was discovered by
\citet{nv2006}, who also estimated its age to $300$-$800$~kyr (the large uncertainty is due
to small size of the two small members --see Table~\ref{lucascavin_members}-- and
unconstrained magnitude of the thermal accelerations in their orbit). A decade later,
\citet{petal2018} found the  three  original members were still the only ones in this cluster.
They also  calculated its
age to be between $500-1000$~kyr using a different method. Assuming the population is
complete, these authors also argued that the estimated sizes of the Lucascavin members,
and the $\simeq 5.79$~hr rotation period, might be enough for 
rotation fission of the parent object to explain their origin
\citep[e.g.,][]{pra2010,petal2019}. The difference with respect to the population of
pairs is that the  assumption that the secondary, escaping from the primary after the
fission event, would split into
two components (namely the two small members (180255) 2003~VM9 and (209570) 2004~XL40).
This possibility was theoretically predicted by \citet{js2011}. If, however, numerous 
smaller fragments are found in the Lucascavin family, this scenario would become
less plausible. Therefore, unlike  our study of other clusters in this paper, the goal of our
analysis here is to ``disprove'' the existence of further fragments in the family. Obviously,
we cannot meet this goal in an absolute manner, but we can set a lower limit on the absolute
magnitude of a putative companion (or, in other words, an upper limit on its size).
\begin{figure}[t]
 \begin{center} 
 \includegraphics[width=0.47\textwidth]{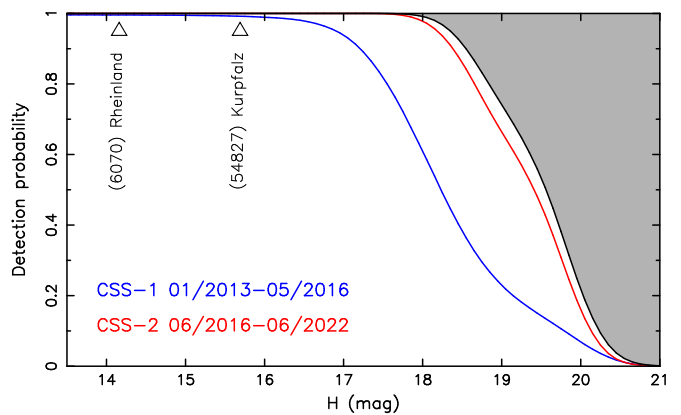}
 \end{center}
 \caption{Detection probability of a companion to Rheinland and Kurpfalz on their
  heliocentric orbit: (i) the blue curve is $p_1$ during the phase~1 of CSS operations, and
  (ii) the red curve is $p_2$ during the phase~2 of CSS operations. The black curve is the
  combined probability $p$ during both phases (Eq.~\ref{comb12}). The gray area allows
  the existence of an additional small body in the system, whose maximum probability of
  occurrence is complementary value to the probability $p$ on the left ordinate.}
 \label{rk_1}
 \end{figure}
\begin{figure*}[t]
 \begin{center} 
  \begin{tabular}{cc} 
   \includegraphics[width=0.47\textwidth]{f16a.eps} &
   \includegraphics[width=0.47\textwidth]{f16b.eps} \\
  \end{tabular} 
 \end{center}
 \caption{Best-fit solution of the complete Datura population to the magnitude
  limit $H_2=20$ and its comparison to the observed population. Left panel: the single
  power law ${\cal M}1$ model with the free parameter representing the slope $\gamma$.
  The best fit value is $\gamma=0.70$, and the corresponding $\chi^2_{\rm min}=13.36$.
  Right panel: The broken power-law model ${\cal M}2$
  with three adjustable parameters $(H_{\rm break},\gamma_1,\gamma_2)$. The best-fit
  values (red star in Fig.~\ref{fd_res_1}) are: $H_{\rm break}=19.13$, $\gamma_1=0.75$, and
  $\gamma_2=0.31$, and the corresponding $\chi^2_{\rm min}=3.85$. The green symbols are the
  currently known population of Datura family members from all surveys,
  the open black circles are the members detected during the phase~2 of CSS ($\{H_i^{\rm o}\}$).
  The red line is the complete model ($\{H_i^{\rm s}\}$), the blue line is the biased
  model ($\{H_i^{\rm b}\}$; the solid part of the blue line has $N_{\rm CSS}'$ objects, the
  same as the number of detected objects beyond the branching magnitude $H_j^{\rm o}$, the
  dotted part is the continuation of the biased population not used for the least-squares
  fitting in Eq.~(\ref{chi2})). The upper abscissa shows an estimate of the size for the
  geometric albedo value $p_V=0.24$.}
 \label{fd_res_2}
 \end{figure*}
  
Moving towards that goal, we note that all three known members in the Lucascavin family were
detected during both phases 1 and 2 of the CSS operations (in our notation, we thus
have $N_{\rm obs}=3$ and $N_{\rm CSS}=3$).%
\footnote{The smaller members, (180255) 2003~VM9 and (209570) 2004~XL40, were detected
 only 1 and 4 times during the phase~1, though.}
In order to use as much information as possible, we have combined data from both phases
of the CSS operations. Given their different performance, we consider both phases as
independent (and uncorrelated) sources of information. Denoting then the detection probability
during the phase~1 by $p_1$, and similarly the detection probability during the phase~2 by $p_2$,
the combined total detection probability $p$ during both phases is
\begin{equation}
 p = 1- (1-p_1)(1-p_2)\, . \label{comb12}
\end{equation}
Note that we first characterized the non-detection during both phases (the second term),
and then take the complement to unity, which expresses detection in at least one of the CSS
phases. Results are shown in Fig.~\ref{ls_1}.

We first briefly comment on the behavior of $p_1$ and $p_2$ (the blue and red curves). The
interesting, and at the first sight puzzling, feature of $p_1$ is that it does not reach
a value of $1$ even for rather bright objects (its maximum value is only about $0.9$).
This is not a mistake, but the result of the Lucascavin cluster’s semimajor axis.
The synodic period of its motion with respect to an Earth observer is in an approximate 7:5
resonance over a year. As a result, for a survey spanning only a short period of time
(such as little more than 3 yr of our CSS phase~1 operations), it may happen that 
Lucascavin objects with certain values of mean longitude in orbit $\lambda$ never occur in
the field-of-view (reasonable solar elongations on the night sky). Since this is a purely
geometrical effect, it affects the detection probability of even very bright objects 
\citep[see, e.g.,][Fig.~2 for illustration of this effect for near-Earth object
characterization]{tric2017}. As the duration of the survey extends, this effect minimizes
and even disappears. As a result, $p_2$ in the 6 yr interval of  CSS phase~2 (red curve)
does not suffer from this problem. The overall detection probability $p_1$ is smaller
than $p_2$, but both reach $p_1\simeq p_2\simeq 0$ at similar $H\simeq 20.5$. This outcome
is because the apparent magnitude detection limit is similar for both phases.

Following the trend of the black curve of Fig.~\ref{ls_1}, $p(H)$, we note that
$p(H)\simeq 1$ up to $H\simeq 18.3$. Therefore the Lucascavin population is complete to
this magnitude limit. This calculation is a conservative estimate because observations of other
surveys may push this limit to higher values. The limit is about one magnitude larger
than that of the two small members in the Lucascavin
family ($\simeq 17.25$). Our result may be therefore interpreted in two ways: either (i) it sets a
constraint on Lucascavin family magnitude distribution, or (ii) it starts tracing the
population void beyond the known set. The former case would imply at least a magnitude gap
between the third and the fourth largest members in the family (this is not impossible,
see, e.g., Fig.~\ref{fd_2}). The latter case may support the idea that the Lucascavin family
formed by rotation fission, with the secondary disrupting into two pieces.
\smallskip
\begin{figure*}[t]
 \begin{center} 
  \includegraphics[width=0.75\textwidth]{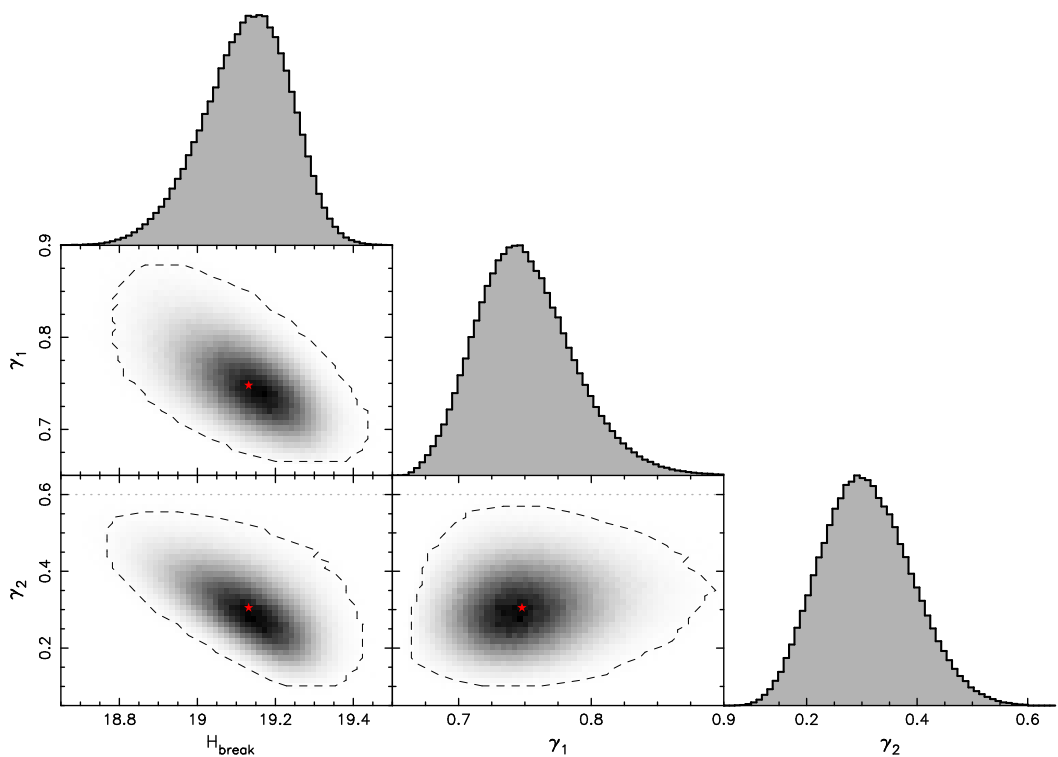}
 \end{center}
 \caption{Admissible parameter values of the ${\cal M}2$ broken power-law model 
  approximating Datura family absolute magnitude distribution in the range $H\in (H_1,H_2)
  =(18.09,20)$. There are three solved-for parameters in the model: (i) the break-point
  magnitude $H_{\rm break}$, (ii) the power-law exponent $\gamma_1$ for $H\leq H_{\rm break}$,
  and (iii) the power-law exponent $\gamma_2$ for $H\geq H_{\rm break}$. Each of the panels
  shows a projection of the solution onto different 2D subspaces
  of the 3D space of parameters $(H_{\rm break},\gamma_1,\gamma_2)$. The dashed line
  delimits 99\% confidence limit zone of the solution, and the gray-scale is proportional
  to the probability density distribution of the solutions. The best-fit parameter
  combination is shown by red star symbol. The gray histograms are simply 1D probability
  density distributions for each of the parameters on the abscissa. The dotted line at
  $\gamma_1$ or $\gamma_2$ values of $0.6$ is shown for reference.}
 \label{fd_res_1}
 \end{figure*}

\noindent{\it Rheinland and Kurpfalz.-- }The pair of asteroids composed of a primary (6070)
Rheinland and a secondary (54827) Kurpfalz is the best studied archetype in its class. This
is because the two asteroids are rather large, namely the $D_1\simeq 4.4\pm 0.6$~km size primary
and the $D_2\simeq 2.2\pm 0.3$~km size secondary (absolute magnitudes $H_1=14.17\pm 0.07$ and
$H_2=15.69\pm 0.04$), and reside in the inner part of the asteroid belt. Their discoveries in
1991 and 2001, and prediscovery data extending to 1950 and 1991, imply a wealth of astrometric
observations allowing accurate orbit determination. This has been noticed already
by \citet{vn2008}, who used this pair to demonstrate they could reach full convergence
in Cartesian space of the two orbits in the past. From this result, they determined the pair 
had an age of $\simeq 17$~kyr.
Later, \citet{rheinAJ2011} and \citet{rheinAJ2017} conducted photometric observations of
both asteroids with the goal to determine their rotation state, including pole orientation,
and shape model. Intriguingly, the spin orientation at the likely moment of their formation has
not been found to be  parallel for the two components, but instead is slightly tilted by about $38^\circ$.
The well confined spin state for both components in this pair allowed them  to pin down the formation
epoch to $16.34\pm 0.04$~kyr \citep[see][]{rheinAJ2017}. An interesting clue about
the formation process, fission of a critically rotating parent body \citep[e.g.,][]{pra2010,petal2019},
is also provided by spectroscopic observations of Rheinland and Kurpfalz: while the first has been
found a typical S-class object, the taxonomy of the latter is either Sq- or even Q-class
\citep[see][]{poli2014}.

Similarly to the case of the Lucascavin family, we aim to determine the magnitude limit
for nonexistence of a putative companion fragment following Rheinland and Kurpfalz
on their heliocentric orbit. Since both Rheinland and Kurpfalz were detected during
CSS phases 1 and 2, we may again combine detection probabilities
$p_1$ and $p_2$ to obtain the total probability $p$ according to the formula (\ref{comb12}).
Results are shown in Fig.~\ref{rk_1}. 

In this case, $p_1$ is comfortably close to
unity even for the fainter component (54827) Kurpfalz.%
\footnote{In fact, (6070) Rheinland has been detected 8 and 26 times during the phases 1 and
 2, while (54827) Kurpfalz has been detected 8 and 21 times the phases 1 and  2.}
However, $p_1$ starts dropping to zero right after $H_2$ of the secondary, such that  limited useful
information would have been reached if we only had the  phase~1 data. Luckily,
the power of the CSS phase 2 observations make extending the final detection
probability $p$ for the orbits in this pair to unity, even near $H\simeq 18$. We may thus
conclude that the available observations rule out a companion fragment of this pair to
this limit, which is $\Delta H\simeq 2.3$ larger than $H_2$ of the secondary. Assuming the
same albedo, the hypothetical companion --if it exists-- must have a size smaller
than $\simeq 10^{-0.2\Delta H}\,D_2\simeq 0.8$~km.

\section{Results} \label{res}
We now proceed towards a more advanced debiasing method than previously used in
the case of the Wasserburg family. The four families introduced in
Sec.~\ref{largevyf} with large-enough known population
of members --Datura, Adelaide, Rampo and Hobson-- will serve us as our testbed cases.

The method, in essence similar to what has been used by \citet{daturaAA2017}, goes as
follows:
\begin{itemize}
\item First, we consider the CSS phase~2 detected sample $\{H_i^{\rm o}\}$ ($i=1,\ldots,
 N_{\rm CSS}$) of the family asteroids and we select a certain member $H_{j}^{\rm o}$ for
 which $p(H_{j}^{\rm o})\simeq 1$ (we call it a ``branching point''). We assume that the
 population is complete up to the absolute magnitude of that   member and becomes incomplete
 for magnitudes larger than
 $H_{j}^{\rm o}$. The cumulative magnitude distribution is therefore represented
 by the observed population until $H_{j}^{\rm o}$, where it has $N_1$ members, and
 then continued with a synthetic (model) population as described below.  We also denote
 the number of family members with magnitudes $\geq H_{j}^{\rm o}$ detected during
 the CSS phase~2 by $N_{\rm CSS}'(\leq N_{\rm CSS})$.
\item Second, we generated the total synthetic population of family members $\{H_i^{\rm s}\}$ having
 absolute magnitudes in between $H_1=H_1^{\rm s}=H_{j}^{\rm o}$ and a certain value $H_2$
 sufficiently larger than $H^{\rm o}_{\rm N_{\rm CSS}}$ with a statistical distribution of the tested 
 magnitude distribution function (we use the sequence of ${\cal M}$ models described below and
 always order the magnitude sequence from the smallest to the largest).
\item Third, we used the detection probability $p(H)$ of the CSS observations to transform the total
 synthetic population to the biased synthetic population $\{H_i^{\rm b}\}$, such that each
 of $\{H_i^{\rm s}\}$ is consulted as to its detectability.  In  practice, for each $H_i^{\rm s}$
 we evaluated $p(H_i^{\rm s})$ and compared it to a uniformly random number $r$ between 0 and 1,
 providing a rationale for detectability or non-detectability: (i) if $r\leq p(H_i^{\rm s})$, the
 asteroid is deemed detected and we record $\{H_i^{\rm s}\}$ in the $\{H_i^{\rm b}\}$ sequence, and
 (ii) if $r> p(H_i^{\rm s})$, the asteroid is deemed not detected and we proceed to the next
 $\{H_i^{\rm s}\}$ value.
\item Fourth, we evaluated a chi-square type target function
 \begin{equation}
  \chi^2=\sum_{i=1}^{N_{\rm CSS}'}\left(\frac{H_i^{\rm b}-H_{j+i-1}^{\rm o}}{\sigma_i}\right)^2 \; ,
   \label{chi2}
 \end{equation}
 comparing the modeled and biased magnitude distribution to the detected set 
 $\{H_i^{\rm o}\}$ by CSS beyond the branching magnitude $H_{j}^{\rm o}$.
\end{itemize}

For sake of simplicity, we (i) use $\sigma_i=0.1$ magnitude for all bodies, and (ii) 
adopt Gaussian statistics to judge the goodness-of-the-fit and set confidence limits
on the adjusted parameters of the model needed to construct the complete (not-biased)
synthetic population $\{H_i^{\rm s}\}$. 
As for the synthetic population, we use the following sequence of power-law models:
\begin{itemize}
\item Model ${\cal M}1$ -- a straight single-slope power law  $N(<H)\propto 10^{\gamma H}$
 with one adjustable parameter $\gamma$ (the absolute normalization for all ${\cal M}$-models
 is set by number $N_1=N(<H_1)$ of family asteroids at $H_1$, because we make sure that the
 population is complete to that limit);
\item Model ${\cal M}2$ -- a broken power-law model with one adjustable break-point at $H_{\rm break}$
 ($H_1 \leq H_{\rm break} \leq H_2$) and two adjustable slope exponents $\gamma_1$ and $\gamma_2$ for $H$ values 
 in the intervals $(H_1,H_{\rm break})$ and $(H_{\rm break},H_2)$ respectively;
\item Model ${\cal M}3$ -- a broken power-law model with two adjustable break-points at $H_{\rm break, 1}$
 and $H_{\rm break, 2}$ ($H_1 \leq H_{\rm break, 1} < H_{\rm break, 2} \leq H_2$) and three adjustable slope
 exponents $\gamma_1$, $\gamma_2$ and $\gamma_3$ for $H$ values in the intervals $(H_1,H_{\rm break, 1})$,
 $(H_{\rm break, 1},H_{\rm break, 2})$ and $(H_{\rm break, 2},H_2)$ respectively;
\end{itemize}
and similarly for ${\cal M}i$ model with $2i-1$ parameters ($i-1$ break-points and $i$ slopes
for the intermediate intervals of $H$). In practice, we limit ourselves to ${\cal M}3$ at maximum in this
paper.

Denote ${\bf p}$ the set of model parameters (e.g., ${\bf p}=(H_{\rm break},\gamma_1,\gamma_2)$ for the
${\cal M}2$ model). Since $\chi^2=\chi^2({\bf p})$ in (\ref{chi2}), the usual goal is to minimize its
value by selecting the best-fit ${\bf p}_\star$ parameter choice. We use a simple Monte Carlo
sampling of ${\bf p}$ space to find these values and to map $\chi^2$ behavior within some zone about
the minimum value $\chi^2_{\rm min}=\chi^2({\bf p}_\star)$. The confidence limits on ${\bf p}$ are
found by choosing a certain domain with a threshold $\chi^2=\chi^2_{\rm min}+\Delta \chi^2$. For
instance, the 99\% confidence limit in one, three and five parametric degrees of freedom in
${\cal M}1$, ${\cal M}2$ and ${\cal M}3$ models corresponds to $\Delta \chi^2=6.63$, $11.3$ and
$15.1$ respectively \citep[e.g.,][]{nr2007}. Similarly the measure of the goodness-of-fit
is judged from the $\chi^2_{\rm min}$ value using the incomplete gamma function as
discussed in \citet{nr2007}.
\smallskip

\noindent{\it Datura.-- }Considering the  data in Fig.~\ref{fd_1} we chose $j=9$ in the case
of the Datura family, namely taking the absolute magnitude $H_{j}^{\rm o}=18.09$ of the ninth
family member as the branching point (i.e., $N_{\rm CSS}'=52$ in this case). We will test the ${\cal M}1$
and ${\cal M}2$ models.%
\footnote{Results discussed in this section do not include (429988) 2013~PZ36 among the family
 members. However, our tests showed that they are robust. By including this body, we observe
 only a statistically insignificant change of the solution, the largest for $\gamma_1 =
 0.70^{+0.15}_{-0.09}$ parameter of the ${\cal M}2$ model.}
\begin{figure}[t]
 \begin{center} 
 \includegraphics[width=0.47\textwidth]{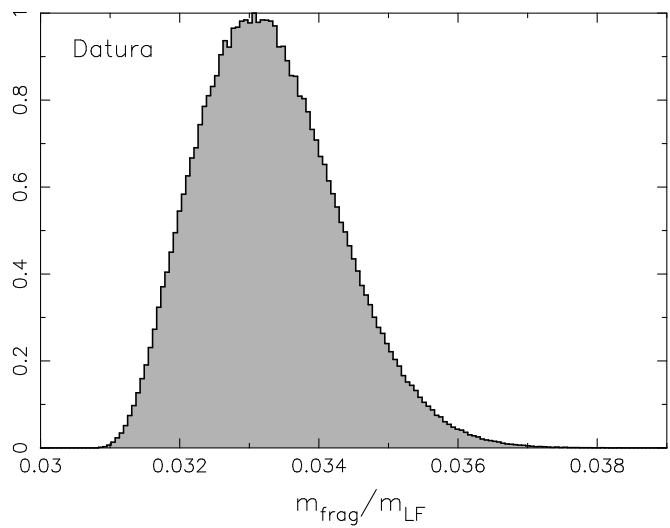}
 \end{center}
 \caption{Probability density distribution of the ratio $m_{\rm frag}/m_{\rm LF}$ 
  (normalized to unit at maximum), where $m_{\rm frag}$ is the mass/volume of all
  members up to absolute magnitude $20$ without (1270) Datura, and $m_{\rm LF}$ is
  the mass/volume of the largest member (1270) Datura. Solution using the broken
  power-law model ${\cal M}2$.}
 \label{fd_res_3}
 \end{figure}
\begin{figure*}[t]
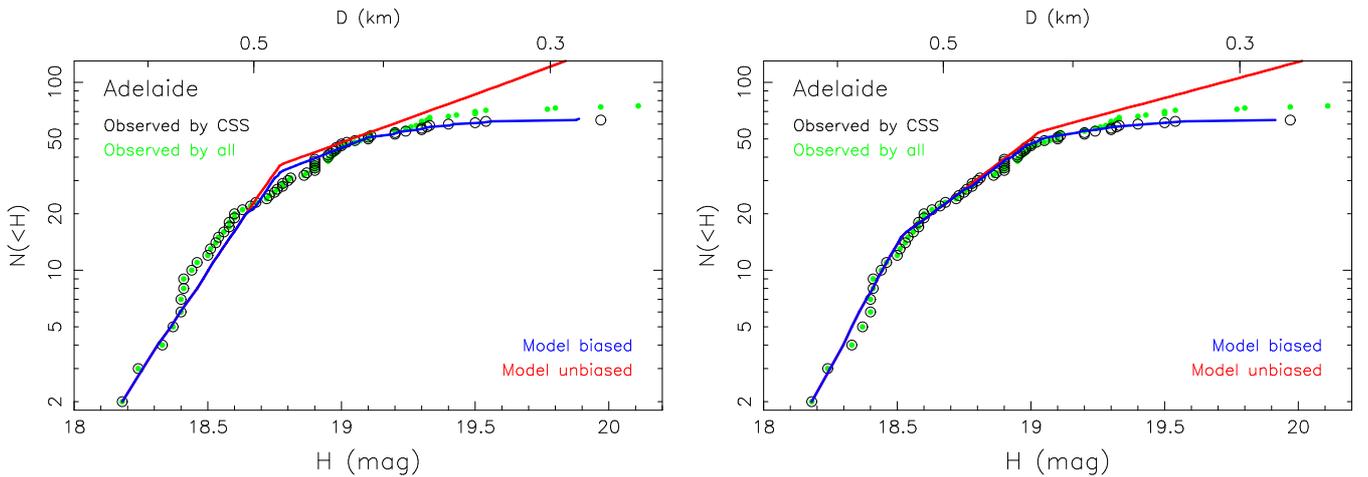

 \begin{center} 
  \begin{tabular}{cc}
  \includegraphics[width=0.47\textwidth]{f19a.eps} &
  \includegraphics[width=0.47\textwidth]{f19b.eps} \\
  \end{tabular}
 \end{center}
 \caption{The best-fit solution of the complete Adelaide population to the magnitude
  limit $H=20$ and its comparison to the observed population. Left panel: the broken power-law model
  ${\cal M}2$ with three adjustable parameters $(H_{\rm break},\gamma_1,\gamma_2)$. The best-fit
  values (red star in Fig.~\ref{fa_res_1}) are: $H_{\rm break}=18.78$, $\gamma_1=2.08$, and
  $\gamma_2=0.47$, and the corresponding $\chi^2_{\rm min}=12.18$. Right panel:  the broken power-law model
  ${\cal M}3$ with five adjustable parameters $(H_{\rm break, 1},H_{\rm break, 2},\gamma_1,\gamma_2,\gamma_3)$.
  The best-fit values are: $H_{\rm break, 1}=18.57$, $H_{\rm break, 2}=
  19.04$, $\gamma_1=2.41$, $\gamma_2=1.00$, and $\gamma_3=0.34$, and the corresponding $\chi^2_{\rm min}=
  3.77$. The green symbols are the
  currently known population of Adelaide family members from all surveys,
  the open black circles are the members detected during the phase~2 of CSS ($\{H_i^{\rm o}\}$).
  The red line is the complete model ($\{H_i^{\rm s}\}$), the blue line is the biased
  model ($\{H_i^{\rm b}\}$; the solid part of the blue line has $N_{\rm CSS}'$ objects, the
  same as the number of detected objects beyond the branching magnitude $H_j^{\rm o}$, the
  dotted part is the continuation of the biased population not used for the least-squares
  fitting in Eq.~(\ref{chi2})).}
 \label{fa_res_2}
 \end{figure*}

In the former case, we find $\gamma = 0.70^{+0.03}_{-0.02}$ (99\% confidence level) and the
best-fit solution having $\chi^2_{\rm min}=13.36$. In the latter case, we find
$H_{\rm break}=19.13^{+0.37}_{-0.48}$, $\gamma_1=0.75^{+0.15}_{-0.09}$, and $\gamma_2=0.31^{+0.30}_{-0.25}$
(99\% confidence level) and the best-fit solution having a significantly improved
$\chi^2_{\rm min}=3.85$ (the improvement for the ${\cal M}3$ model is already statistically
insignificant). 

The best-fit solutions of both models are shown in Fig.~\ref{fd_res_2}.
While formally the minimum $\chi^2_{\rm min}$ values are both statistically justifiable
using the $Q$-function measure \citep{nr2007}, the ${\cal M}1$ performs quite worse
beyond $H\simeq 19.5$. This is because continuing the steep power-law
distribution required by the magnitude distribution of the Datura members between $H = 18$ and
$19$ would keep pushing the detactable population high (given the only
slow decay of the detection probability $p(H)$ from Fig.~\ref{fd_1}). This problem is
remedied by setting a break-point at which the distribution becomes shallower; this
behavior is readily provided by the ${\cal M}2$ model. The upper abscissa on both
panels of Fig.~\ref{fd_res_2} is an estimate of Datura member size using the
geometric albedo value $p_V=0.24$. The break-point magnitude $H_{\rm break}$ solution
within the ${\cal M}2$ model maps onto a $0.3$-$0.5$~km range of sizes.

Figure~\ref{fd_res_1} provides more detailed information on the ${\cal M}2$
model parameter solution. In spite of weak correlations, the solution seems to be
well-behaved. Interestingly, the slope exponents satisfy $\gamma_1 > 0.6$ and
$\gamma_2 < 0.6$. The magnitude slope $\gamma$ translates to an exponent $\alpha=-5\gamma$
of a cumulative size distribution (assuming constant albedo on a given interval of
$H$-values). Therefore the threshold value $0.6$ maps onto a critical size exponent
$-3$: for shallower distributions the mass is dominated by the largest members, while
for steeper distributions the mass is dominated by the smallest fragments. 

In our ${\cal M}2$ solution for Datura members, the mass is dominated by the sizes at the
breakpoint, while in the ${\cal M}1$ solution the fragment mass cannot be well
constrained because it is dominated by the smallest members. Here we use the ${\cal M}2$
solution and estimate the total mass $m_{\rm frag}$ contained in Datura family members 
with absolute magnitudes between $16$ and $20$ from our complete model (i.e., excluding 
(1270) Datura itself). We also normalize $m_{\rm frag}$ by the mass $m_{\rm LF}$ of (1270) Datura.
The statistical distribution of this ratio, as mapped from the 99\% confidence level
parametric region shown in Fig.~\ref{fd_res_1}, is shown in Fig.~\ref{fd_res_3}.
We find $m_{\rm frag}/m_{\rm LF} = 0.033^{+0.005}_{-0.002}$. Unless the cumulative number
of Datura members becomes significantly steeper somewhere beyond the magnitude
limit $20$, which is certainly possible \citep[e.g.,][]{d2007}, we estimate that their collective
mass only represents $\simeq 3.3$\% of the (1270) Datura mass. From this analysis, we suggest the
family may have been formed from a large cratering event.
\smallskip

\noindent{\it Adelaide.-- }The extreme nature of the magnitude distribution in the
Adelaide family (Fig.~\ref{fa_1}) makes us choose $j=2$,
therefore we associate the point to the first member next to (525) Adelaide with
$H_{j}^{\rm o}=18.18$. With that choice we have $N_{\rm CSS}'=62$. In this case, we test
${\cal M}1$, ${\cal M}2$ and ${\cal M}3$ models.

We find that the single power-law model ${\cal M}1$ is incompatible with the family data.
The formally best-fit slope $\gamma\simeq 1.86$ tries to compromise between the extremely
steep part of the magnitude distribution between $18.18$ and $\simeq 18.75$ and much
shallower distribution beyond. However, none of the features is matched well and the
formal $\chi^2_{\rm min}\simeq 235$ has to be statistically rejected. The basic inconsistency
of such a model stems from the behavior of the detection probability $p(H)$ shown in the bottom
panel of Fig.~\ref{fa_1}. In simple words, $p(H)$ is quite smooth and gradual even beyond
$\simeq 19$ magnitude and does not resemble the sharp lack of detected fragments at $\simeq 18.7$
magnitude. In the Adelaide family case, we need some slope change even in the complete population,
and this is provided by models ${\cal M}2$ and ${\cal M}3$.
\begin{figure*}[t]
 \begin{center} 
 \includegraphics[width=0.75\textwidth]{f20.eps}
 \end{center}
 \caption{Admissible parameter values of a ${\cal M}2$ broken power-law model 
  approximating Adelaide family absolute magnitude distribution in the range $H\in (H_1,H_2)
  =(18.2,20)$. There are three solved-for parameters in the model: (i) the break-point
  magnitude $H_{\rm break}$, (ii) the power-law exponent $\gamma_1$ for $H\leq H_{\rm break}$,
  and (iii) the power-law exponent $\gamma_2$ for $H\geq H_{\rm break}$.
  Each of the panels shows a projection of the solution onto different 2D subspaces
  of the 3D space of parameters $(H_{\rm break},\gamma_1,\gamma_2)$. The dashed line
  delimits 99\% confidence limit zone of the solution, and the gray-scale is proportional
  to the probability density distribution of the solutions. The best-fit parameter
  combination is shown by red star symbol. The gray histograms are simply 1D probability
  density distributions for each of the parameters on the abscissa; the dashed histograms along the
  distributions of the $\gamma_1$ and $\gamma_2$ exponents correspond to the first and the last
  exponent in the ${\cal M}3$ broken power-low models with two break-points $(H_{\rm break, 1},H_{\rm break, 2})$
  (i.e., the first in the interval $(18.2,H_{\rm break, 1})$, and the second in the interval
  $(H_{\rm break, 2},20)$). The dotted line at $\gamma_1$ or $\gamma_2$ values of $0.6$ is shown for reference.}
 \label{fa_res_1}
 \end{figure*}
\begin{figure}[t]
 \begin{center} 
 \includegraphics[width=0.47\textwidth]{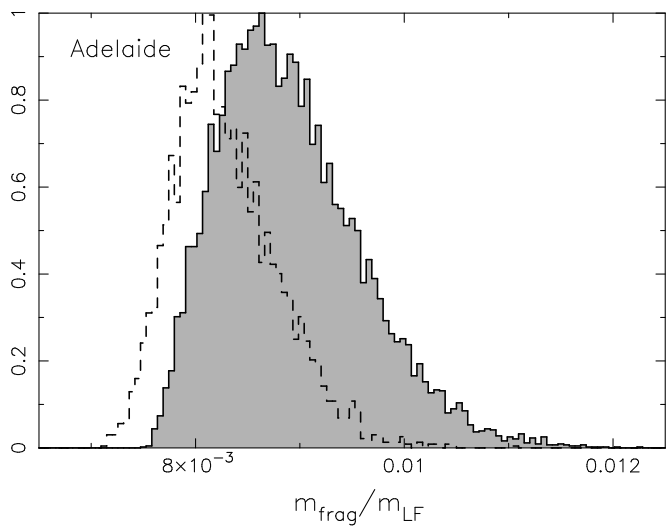}
 \end{center}
 \caption{Probability density distribution of the ratio $m_{\rm frag}/m_{\rm LF}$ 
  (normalized to unit at maximum), where $m_{\rm frag}$ is the mass/volume of all
  members up to absolute magnitude $20$ without (525) Adelaide, and $m_{\rm LF}$ is
  the mass/volume of the largest asteroid (525) Adelaide. The solid/gray histogram
  for the ${\cal M}2$ model, the dashed histogram for the ${\cal M}3$ model (all
  models whose parameters ${\bf p}$ are within the 99\% confidence level of the
  family population were used).}
 \label{fa_res_3}
 \end{figure}

In the former case, we find $H_{\rm break}=18.78^{+0.14}_{-0.26}$, $\gamma_1=2.08^{+0.92}_{-0.19}$, and
$\gamma_2=0.47^{+0.30}_{-0.24}$ (99\% confidence level). The best-fit solution has $\chi^2_{\rm min}
=12.18$. The model reflects a slope change from steep to shallow  near $18.75$. The minimum of $\chi^2$
 reached in the ${\cal M}2$ model is fully acceptable, 
yet the left panel on Fig.~\ref{fa_res_2} indicates the solution may still be improved (obviously at the
expense of more parameters). This is provided by the ${\cal M}3$ model (the right panel on 
Fig.~\ref{fa_res_2}), which has $H_{\rm break, 1}=18.57^{+0.29}_{-0.26}$, $H_{\rm break, 2}=
19.04^{+1.00}_{-0.14}$, $\gamma_1=2.41^{+0.67}_{-0.48}$, $\gamma_2=1.00^{+0.77}_{-0.59}$, and $\gamma_3=
0.34^{+0.44}_{-0.28}$ (99\% confidence level) and has $\chi^2_{\rm min}=3.77$. 

Figure~\ref{fa_res_1}
shows 2D projections of the ${\cal M}2$ model parameters, resembling those for Datura family in
Fig.~\ref{fd_res_1}, except for $\gamma_1$ value significantly steeper. The ${\cal M}3$ model parameters
are more correlated within each other, as many combinations for positions of the two breakpoints
$H_{\rm break, 1}$ and $H_{\rm break, 2}$ and the intermediate slopes $\gamma_1$ and $\gamma_2$ , are
possible. Obviously, the solution of the faintest-slope $\gamma_3$ is consistently shallow, even shallower
than $\gamma_2$ in model ${\cal M}2$ (see Fig.~\ref{fa_res_1}). 

There is a robust, common result
following from the ${\cal M}2$ and ${\cal M}3$ models: (i) the initial slope parameter in
the $18.2$-$18.6$ absolute magnitude range must be very steep (i.e., $2$-$3$, and (ii)
the final slope beyond absolute magnitude $19$ must be rather shallow (i.e., $0.1$-$0.7$).
Given the shallow magnitude distribution at the limit of very small Adelaide members (for most
part $<0.6$), the bias-corrected fragment population mass is dominated by $H\simeq 19$ Adelaide
members. We can thus repeat the computation performed for the Datura family, and compute
the ratio $m_{\rm frag}/m_{\rm LF}$ of the Adelaide members with $H>18$ ($m_{\rm frag}$) and the
mass of the largest asteroid (525)~Adelaide itself ($m_{\rm LF}$). Obviously, we carry out this
computation for the bias-corrected populations of the ${\cal M}2$ and ${\cal M}3$ models, rather
than the observed population of the Adelaide family members. 

The results, shown in Fig.~\ref{fa_res_3}, provide tight constraints on the complete population
of the Adelaide members in the $18$-$20$ magnitude range: $m_{\rm frag}/m_{\rm LF} =
0.0088^{+0.0013}_{-0.0009}$ for the ${\cal M}2$ model, and $m_{\rm frag}/m_{\rm LF} =
0.0084^{+0.0003}_{-0.0005}$
for the ${\cal M}3$ model. If these estimates hold also for the population at the family origin
(see Sec.~\ref{concl} for an alternative option), the Adelaide family is an exemplary case of
a large cratering event. We estimated the size of the expected
crater on (525)~Adelaide in Sec.~\ref{concl}.
\smallskip
\begin{figure*}[t]
 \begin{center}
  \begin{tabular}{cc} 
   \includegraphics[width=0.47\textwidth]{f22a.eps} &
   \includegraphics[width=0.47\textwidth]{f22b.eps} \\
  \end{tabular}
 \end{center}
 \caption{The best-fit solution of the complete Rampo population to the magnitude
  limit $H_2=20$ and its comparison to the observed population. Left panel: the single
  power law ${\cal M}1$ model with the free parameter representing the slope $\gamma$.
  The best fit value is $\gamma\simeq 1.17$, and the corresponding $\chi^2_{\rm min}=17.6$.
  Right panel: The broken power-law model ${\cal M}2$
  with three adjustable parameters $(H_{\rm break},\gamma_1,\gamma_2)$. The best-fit
  values (red star in Fig.~\ref{fr_res_2}) are: $H_{\rm break}=18.47$, $\gamma_1=1.72$, and
  $\gamma_2=0.51$, and the corresponding $\chi^2_{\rm min}=1.25$. The green symbols are the
  currently known population of Rampo family members from all surveys,
  the open black circles are the members detected during the phase~2 of CSS ($\{H_i^{\rm o}\}$).
  The red line is the complete model ($\{H_i^{\rm s}\}$), the blue line is the biased
  model ($\{H_i^{\rm b}\}$; the solid part of the blue line has $N_{\rm CSS}'$ objects, the
  same as the number of detected objects beyond the branching magnitude $H_j^{\rm o}$, the
  dotted part is the continuation of the biased population not used for the least-squares
  fitting in Eq.~(\ref{chi2})). The upper abscissa shows an estimate of the size for the
  geometric albedo value $p_V=0.24$.}
 \label{fr_res_1}
 \end{figure*}

\noindent{\it Rampo.-- }In this case, we use $j=4$, corresponding to a $H_{j}^{\rm o}=18.09$ magnitude
branching point (Fig.~\ref{fr_2}). Using that choice we have $N_{\rm CSS}'=22$, slightly less data
than for the Datura and Adelaide families. We tested the ${\cal M}1$ and ${\cal M}2$ models in this situation.
\begin{figure*}[t]
 \begin{center} 
 \includegraphics[width=0.75\textwidth]{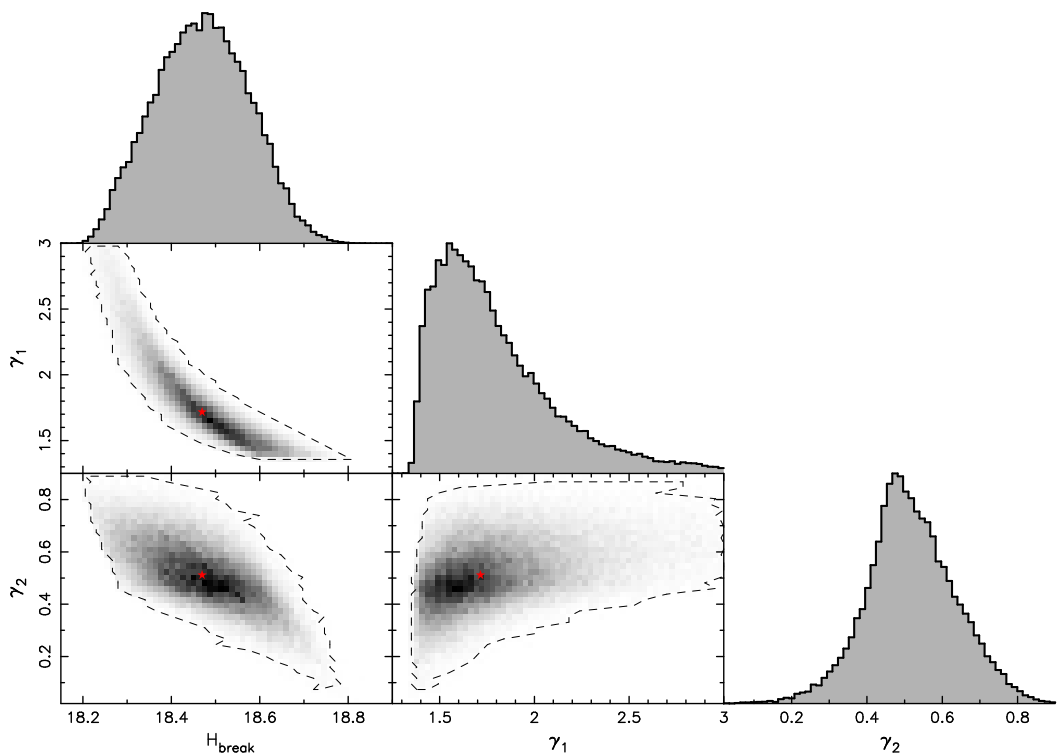}
 \end{center}
 \caption{Admissible parameter values of the ${\cal M}2$ broken power-law model 
  approximating Rampo family absolute magnitude distribution in the range $H\in (H_1,H_2)
  =(18.1,20)$. There are three solved-for parameters in the model: (i) the break-point
  magnitude $H_{\rm break}$, (ii) the power-law exponent $\gamma_1$ for $H\leq H_{\rm break}$,
  and (iii) the power-law exponent $\gamma_2$ for $H\geq H_{\rm break}$. Each of the panels
  shows a projection of the solution onto different 2D subspaces
  of the 3D space of parameters $(H_{\rm break},\gamma_1,\gamma_2)$. The dashed line
  delimits 99\% confidence limit zone of the solution, and the gray-scale is proportional
  to the probability density distribution of the solutions. The best-fit parameter
  combination is shown by red star symbol. The gray histograms are simply 1D probability
  density distributions for each of the parameters on the abscissa. The dotted line at
  $\gamma_1$ or $\gamma_2$ values of $0.6$ is shown for reference.}
 \label{fr_res_2}
 \end{figure*}

The best-fit with a single power-law ${\cal M}1$ model only reaches $\chi^2_{\rm min}=31.4$ (with the
median slope parameter $\gamma\simeq 1.44$). Given $N_{\rm CSS}'=22$ data points, this solution is
statistically unacceptable \citep[the quality factor $Q\simeq 0.067$; see][]{nr2007}.
Figure~\ref{fr_res_1} illustrates the problem in a graphical way, namely the predicted population of
fragments beyond magnitude $19$ (blue dashed line on the left panel) becomes steep and incompatible
with the single Rampo fragment detected by CSS. Things improve if the magnitude
of the power-law model ${\cal M}1$ is shifted to $H_{j}^{\rm o}=18.0$ (still within the assumed
$0.1$ magnitude uncertainty). This helps to straighten the sequence of 
observed members immediately after $H_{j}^{\rm o}$, where the detection probability is still close
to 1. With that change, the single power-law model ${\cal M}1$ provides best match
with $\chi^2_{\rm min}=17.60$ (and, obviously, smaller slope $\gamma\simeq 1.17$). While not impressive,
the solution is formally acceptable, but it suffers from the same problem in matching the faint end
of the observed Rampo population using CSS.

The broken power-law model ${\cal M}2$ performs much better in this circumstance. It reaches $\chi^2_{\rm min}=1.52$,
and the parameter solution $H_{\rm break}=18.47^{+0.37}_{-0.28}$, $\gamma_1=1.72^{+1.28}_{-0.40}$, and
$\gamma_2=0.51^{+0.39}_{-0.49}$ (99\% confidence level; see Fig.~\ref{fr_res_2}). The overall median slope
$\simeq 1.17$-$1.44$ is
thus traded for a steeper leg initially, followed with a shallower part beyond $H_{\rm break}$. The
small $\chi^2_{\rm min}$ conforms to the visually perfect match shown on the right panel of
Fig.~\ref{fr_res_1}.

\begin{figure}[t]
 \begin{center} 
 \includegraphics[width=0.47\textwidth]{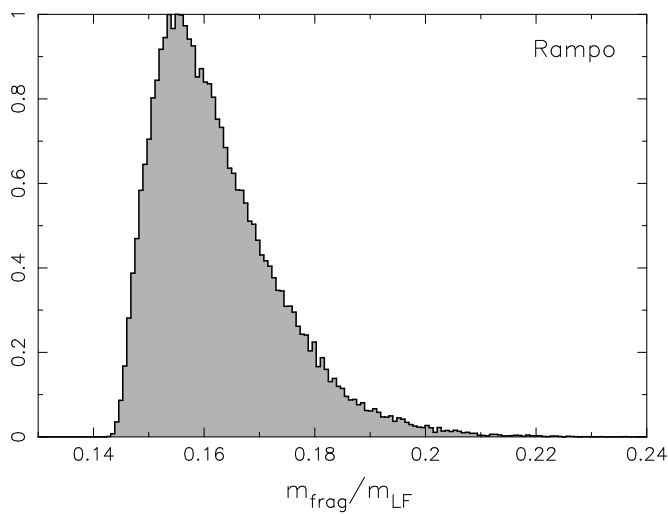}
 \end{center}
 \caption{Probability density distribution of the ratio $m_{\rm frag}/m_{\rm LF}$ 
  (normalized to unit at maximum), where $m_{\rm frag}$ is the mass/volume of all
  members up to absolute magnitude $20$ without (10321) Rampo, and $m_{\rm LF}$ is
  the mass/volume of the largest asteroid (10321) Rampo. Solution using the broken
  power-law model ${\cal M}2$.}
 \label{fr_res_3}
 \end{figure}

Figure~\ref{fr_res_3} shows the model predicted mass in the Rampo members between magnitudes
$18$ and $\simeq 19.5$, namely $m_{\rm frag}/m_{\rm LF} = 0.16^{+0.08}_{-0.02}$. However, since
the $\gamma_2$ slope beyond $H_{\rm break}$ tends to be steep (with value larger than $0.6$
not excluded), the real fragment mass with respect to (10321) Rampo may be even larger. In any
case, out of the three families analyzed so far, the Rampo family represents the most energetic
collisional event.
\smallskip
\begin{figure}[t]
 \begin{center} 
 \includegraphics[width=0.47\textwidth]{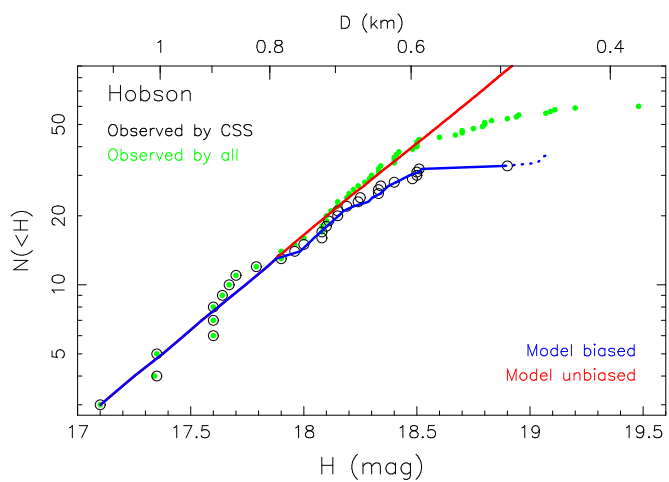}
 \end{center}
 \caption{The best-fit solution of the complete Hobson population to the magnitude
  limit $H_2=19$ and its comparison to the observed population. The single
  power law ${\cal M}1$ model with the free parameter representing the slope $\gamma$.
  The best fit value is $\gamma=0.81$, and the corresponding $\chi^2_{\rm min}=6.31$.
  The green symbols are the currently known population of Hobson family members from all surveys,
  the open black circles are the members detected during the phase~2 of CSS ($\{H_i^{\rm o}\}$).
  The red line is the complete model ($\{H_i^{\rm s}\}$), the blue line is the biased
  model ($\{H_i^{\rm b}\}$; the solid part of the blue line has $N_{\rm CSS}'$ objects, the
  same as the number of detected objects beyond the branching magnitude $H_j^{\rm o}$, the
  dotted part is the continuation of the biased population not used for the least-squares
  fitting in Eq.~(\ref{chi2})). The upper abscissa shows an estimate of the size for the
  geometric albedo value $p_V=0.2$.}
 \label{fh_res_1}
 \end{figure}

\noindent{\it Hobson.-- }The record of observed Hobson members, both the total count
and the subset detected by CSS, is comparable to the Rampo family. However, because of
Hobson's larger heliocentric distance, and its larger eccentricity, the predicted detection
probability by CSS is shifted by nearly a magnitude towards small $H$ values (see
Figs.~\ref{fh_1} and \ref{fr_2}). This allows us to conduct the bias-correction on a
shifted segment of Hobson member magnitudes/sizes if compared to Rampo, which explains
the differences in results.

In this case, we use $j=3$, corresponding to the $H_{j}^{\rm o}=17.10$ magnitude
branching point (Fig.~\ref{fh_1}). Using that choice, we have $N_{\rm CSS}'=31$.
We tested the ${\cal M}1$ and ${\cal M}2$ models in this situation.

Given the aforementioned difference in detection probabilities for the Rampo and Hobson
families, the ${\cal M}1$ model is currently sufficient to match the Hobson population
between $\simeq 17$ and $\simeq 19$ magnitudes (Fig.~\ref{fh_res_1}). The best-fit
simulation reaches $\chi^2_{\rm min}=6.31$, while the simulations using the ${\cal M}2$
model were able to improve this value to $\chi^2_{\rm min}=5.55$. This is not enough of 
a statistically
significant difference to justify the necessity of a broken power-law model for the
Hobson population of members; the simple power-law model performs just as well. The
slope parameter is $\gamma = 0.81^{+0.03}_{-0.02}$ (99\% confidence level). Because this
value is larger than $0.6$, we cannot estimate the mass contained in the fragment population,
(as the smallest asteroids still dominate the mass). We can only set a lower limit
from the population available to us, and this gives $m_{\rm frag}/m_{\rm LF} \geq 0.6$.
In this case, $m_{\rm LF}$ contains the mass of the two largest asteroids, (18777) Hobson and
(57738) 2001~UZ160. Clearly, the Hobson family results from the catastrophic disruption of
a parent body.

\section{Discussion and conclusions} \label{concl}
Our work provides evidence for a break in the magnitude distribution in several of the
very young families analyzed here. Before considering  implications, however, we
first must attempt to further justify the result and understand its meaning.
We can think of at  least two conventional reasons for what we see.
\smallskip

\noindent{\it Missing halo of small members?-- }The first possibility is that we were unable 
to identify small family members beyond $H_{\rm break}$. Their deficit, quantitatively
as shown by the shallow slope at faint magnitudes, may not be real, but instead represents a failure in our the
clustering association. Perhaps, many of these small fragments were ejected with larger velocities
and drifted farther from the core of the family. This scenario is a plausible situation
for larger and older families in the main belt, which are identified in 3D proper element space
by their large spatial densities of asteroids compared to the background population 
\citep[see discussion in][]{netal2015}. For the very young families, however, 
clusters in the 5D space of osculating orbital elements, with additional tracers such as
the correlated values of the secular angles $\Omega$ and $\varpi$ (Sec.~\ref{yf}), help to
minimize the problem of missing members (if identified in our catalogs). The nominal family-identification
method, described in the Appendix, uses a very conservative search zone (followed by a control
on the past convergence of the orbits). In order to demonstrate the margin we allow, we present
a more in-depth test in the case of the Adelaide family here.

We use four nested boxes around the asteroid (525) Adelaide in osculating orbital elements
(data from MPC catalog as of May~15, 2023), with the following parameters:
\begin{itemize}
\item {\it Box 1} defined by the following differences in semimajor axis $a$, eccentricity $e$,
 inclination $I$, longitude of node $\Omega$, and argument of perihelion $\omega$:
 $(\delta a,\delta e,\delta I,\delta \Omega,\delta \omega)=(\pm 0.01,\pm 0.01,\pm 0.1^\circ,
 10^\circ,10^\circ)$;
\item {\it Box 2} defined by the following differences in semimajor axis $a$, eccentricity $e$,
 inclination $I$, longitude of node $\Omega$, and argument of perihelion $\omega$:
 $(\delta a,\delta e,\delta I,\delta \Omega,\delta \omega)=(\pm 0.02,\pm 0.02,\pm 0.15^\circ,
 20^\circ,20^\circ)$;
\item {\it Box 3} defined by the following differences in semimajor axis $a$, eccentricity $e$,
 inclination $I$, longitude of node $\Omega$, and argument of perihelion $\omega$:
 $(\delta a,\delta e,\delta I,\delta \Omega,\delta \omega)=(\pm 0.03,\pm 0.03,\pm 0.2^\circ,
 30^\circ,30^\circ)$;
\item {\it Box 4} defined by the following differences in semimajor axis $a$, eccentricity $e$,
 inclination $I$, longitude of node $\Omega$, and argument of perihelion $\omega$:
 $(\delta a,\delta e,\delta I,\delta \Omega,\delta \omega)=(\pm 0.035,\pm 0.035,\pm 0.25^\circ,
 35^\circ,35^\circ)$.
\end{itemize}
Our nominal procedure described in the Appendix uses {\it Box 3}, where we identified all
$79$ Adelaide family members listed in Table~\ref{adelaide_members}. Here are the data of
asteroid populations found in the subsequent boxes: (i) {\it Box 1} contains 74 asteroids, all
Adelaide family members and no background objects, (ii) {\it Box 2} contains 84 asteroids,
all 79 Adelaide family members and 5 background objects, (iii) {\it Box 3} contains 105 asteroids,
all 79 Adelaide family members and 26 background objects, and (iv) {\it Box 4} contains 135 asteroids,
all 79 Adelaide family members and 56 background objects. The identified members of the Adelaide
family reside in the interior two boxes (for most part already in {\it Box 1}). The background
population of asteroids slowly ramps from the {\it Box 2} stage.%
\footnote{Taken very naively, namely just multiplying dimensions of the box in all searched orbital
 elements, the "volume" of the {\it Box 4} is $\simeq 2.3$ larger than that of the {\it Box  3}.
 The number of background asteroids increased by a factor of $56/26\simeq 2.15$. This may indicate
 roughly uniform, but very sparse, population of background population at the location of the
 Adelaide family.}
These statistics make us
believe that we are not missing any distant (and small) Adelaide members. A similar situation
applies to other families as well.
\smallskip

\noindent{\it Collisional comminution of family members beyond $H_{\rm break}$?-- }The
bias-corrected population of the family members, as follows from our analysis, tells us
about the current population several hundreds of  thousands of years after the origin of the
clusters. This  population may have experienced some degree of collisional evolution over that 
interval, enough to disrupt some family members.  As a result, we must verify
whether the transition to a shallower magnitude distribution beyond $H_{\rm break}$ in the case
of the Datura, Adelaide and Rampo families is not simply produced by collisional comminution.

We note that the size distribution of the main belt becomes shallow below 1~km
in diameter, and its equivalent steepness at $\simeq 500$~m may be as small as $\gamma \simeq 0.3$
\citep[see Fig.~1 in][]{bottke2020}. Any submerged population introduced into this vast
population of projectiles, such as a volume-limited new family, tends to equilibrate with
the background (assuming disruption laws are the same for the background and family objects).
The crucial issue with  young asteroid families is the timescale
of this process: has enough time passed since the origin of the family to reach equilibrium 
for members that are hundreds of meters?

In order to explore this issue we performed the following numerical experiment. We used the
well-tested Monte Carlo code {\tt Boulder} \citep[e.g.,][]{m2009,ver2018} 
to track the collisional evolution of multiple small-body populations. Here we 
simulated  both the internal impact/cratering/disruption processes within each of the populations
and also the  mutual collisional interaction of the populations (i.e., objects in one
population may serve as impactors for the other and vice versa). The code version we adopted models
the size-frequency distribution for each of the populations but does not include the orbital dynamics of the
population members. 

We used two populations: (i) the background population of main belt
asteroids taken from \citet{betal2015}, and (ii) the young family population. We were only interested
in a brief interval of time lasting $\leq 1$~Myr (i.e., equal to the estimated age for
the corresponding family).The origin of the simulation was the formation epoch of the family.
The main belt population is effectively in equilibrium for the relevant sizes of about ten meters and
larger, but the family population is expected to evolve with time; proving or disproving
changes of the family size distribution at hundred meter and larger sizes was the goal of our simple
test. 

The initial size distribution of the family was equal to the best-fitting, bias-corrected solution
from Sec.~\ref{res} with the following modification: we disregarded the breakpoint at $H_{\rm break}$
in the ${\cal M}2$ (and higher) class of solutions and continued the distribution with the
power-exponent $\gamma_1$ from the first magnitude interval $(H_1,H_{\rm break})$. We considered
$0.24$ geometric albedo to convert absolute magnitude in Sec.~\ref{res} to sizes.  Finally, we needed 
to specify parameters
of the collisional interaction -- intrinsic collisional probability $p_i$ and mean relative velocity
${\bar v}$ at impact-- within each of the populations and across them. This was done as follows.

The intrinsic values of $p_i$ and ${\bar v}$ of the main belt population have been evaluated
in many previous studies, and there is some small variation among them (related mostly to the
smallest-size bodies used for their determination). We used $p_i=2.9\times 10^{-18}$ km$^{-2}$~yr$^{-1}$
and ${\bar v}=5.3$ km~s$^{-1}$ \citep[see Sec.~2.1 in][]{betal2015}. For simplicity, the  same values 
were taken for main belt projectiles impacting the young asteroid family population. The latter was deemed
to be negligible in the relevant sizes of ten meters and larger (see Fig.~\ref{datura_sfd}), which allowed
us to neglect family members as a meaningful  population of impactors for  main belt asteroids. 

The
tricky part of the calculation was to determine the intrinsic collisional parameters for the family population.
This is because $p_i$ and ${\bar v}$ depend on the orbital architecture of the family population,
which experienced strong evolution immediately after the family formation event.  The initially extremely
compact cloud of fragments should first disperse in orbital mean anomaly (over a characteristic timescale of
few thousands of years), and subsequently continues to disperse in longitude of node and perihelion
(reaching about $20^\circ$ interval for a $\simeq 500$~kyr old Datura family, e.g., Fig~\ref{fd_1}).
This highly dynamical situation implies that the intrinsic family values of $p_i$ and ${\bar v}$ are
also strongly time dependent. Importantly, because of the initial orbital similarity, the collision probabilities
may also be very high. 

Since assumptions of the most commonly used scheme to evaluate $p_i$ and ${\bar v}$,
notably the \"Opik-Wetherill approach are not satisfied \citep[see][]{o1951,w1967,g1982}, we used a more
direct approach based on a numerical orbital integration of a finite sample of $n$ bodies in the population
\citep[for details of the approach see, e.g.,][]{metal1996,d1998}. Monitoring the orbits over a time
interval $\Delta T$,
we recorded all mutual close encounters at a small-enough distance $R$ (in our simulations we used
$R$ up to $0.002$~au). The available number of pair combinations is
$n_{\rm pair} = n(n-1)/2$. If $N$ such encounters are found, we have an estimate
\begin{equation}
 p_i \simeq {N\over n_{\rm pair}R^2\Delta T}\; . \label{picomp}
\end{equation}
Ideally, one should evaluate the whole congruence of encounters by varying the threshold distance $R$
and verify that $N(R)\propto R^2$, such that $p_i$ converges to a constant value. We verified
this behavior is satisfied in our experiment. More importantly, as the orbits in the family undergo their 
dynamical evolution, we find that $p_i$ changes as a function of time.
\begin{figure}[t]
 \begin{center} 
 \includegraphics[width=0.47\textwidth]{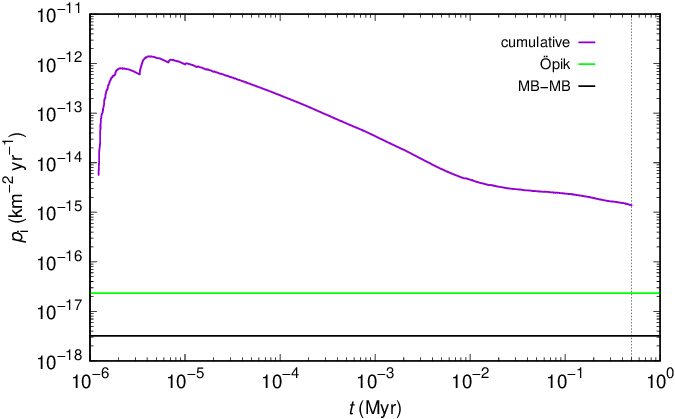}
 \end{center}
 \caption{Intrinsic collisional probability ${\bar p}_i$ of Datura members with respect to each
  other based on data from numerical integration of $n=57$ asteroids of a synthetic family and using
  Eq.~(\ref{picomp}). Time since origin of the family at the abscissa also represents the $\Delta T$
  timescale in the denominator of Eq.~(\ref{picomp}). The initially extremely compact configuration of
  Datura fragments results in very large ${\bar p}_i$ values during the first few revolutions about
  the Sun. This is followed by a decline reflecting asteroid dispersal which has two phases: (i) first
  along the elliptic orbit, completed in $\simeq (2-3)$~kyr (corresponding to change in slope of
  ${\bar p}_i(t)$ power-law approximation), (ii) followed with a phase during which the secular angles
  (longitude of node and perihelion) drift from each other. This latter phase has not been completed
  yet for very young families (Fig.~\ref{fd_1}). For this reason the terminal ${\bar p}_i$ value is nearly
  two orders of magnitude larger than the formal collision probability computed with the \"Opik-Wetherill
  approach (orange line), in which only the values of $(a,e,I)$ would be taken into consideration and
  the secular angles $(\Omega,\omega)$ considered uniformly distributed in the whole interval $(0^\circ,
  360^\circ)$. The reference mean value of the intrinsic collisional
  probability for the main belt population, $p_i\simeq 3\times 10^{-18}$ km$^{-2}$~yr$^{-1}$, is shown
  by the gray horizontal line for reference.}
 \label{datura_pi}
 \end{figure}

We considered the case of the Datura family as an exemplary case for our  method. In order to track the 
characteristic orbital evolution of Datura members, we created a synthetic Datura family consisting
of its $57$ largest members (Table~\ref{datura_members}). The initial configuration was created
by propagating Datura's orbit backward in time until the argument of perihelion was $\omega\simeq 0^\circ$
and true anomaly $f\simeq 150^\circ$. We assumed an isotropic and size-dependent velocity ejection
field
\begin{equation}
 V(D) = 1\,{\rm m}\,{\rm s}^{-1} \left({D\over 2\,{\rm km}}\right)^{-0.5}\; , \label{datvel}
\end{equation}
which allows us to create a configuration that, in the $(a,e)$ and $(a,i)$ planes, resembles the distribution
of Datura members \citep[e.g.,][]{nv2006,daturaAA2017} (for instance
semimajor axes spread $\pm 0.001$~au). We used symplectic integrator {\tt rmvs3}, part of a well-tested package
{\tt swift},%
\footnote{\url{http://www.boulder.swri.edu/~hal/swift.html}}
and included perturbations from eight planets
and the massive dwarf planets Ceres and Vesta. We also randomly assigned thermal accelerations (i.e., the Yarkovsky
effect) to the family members in the transverse direction. The smallest members in our simulation were
thus given semimajor axis
drift rates up to $da/dt\simeq \pm 0.0006$ au~Myr$^{-1}$. We determined mutual distances of all simulated
particles at every timestep of $3.6525$~days, seeking very close encounters for determination of
$p_i$ from Eq.~(\ref{picomp}) (determination of the encounter configurations was implemented on-line by
seeking minima on the memory-sorted mutual distances), and propagated the synthetic family for a timespan
of $500$~kyr corresponding to the Datura age \citep[e.g.,][]{daturaSci2006,daturaAA2009,daturaAA2017}.

We evaluated a ''cumulative'' ${\bar p}_i$ value by taking $\Delta T$ in Eq.~(\ref{picomp}) the current epoch in
the integrated system and counting $N$ from all encounters until that moment. The results are shown in
Fig.~\ref{datura_pi}. We find that ${\bar p}_i$ peaks at $\simeq 10$~yr, representing three revolutions about the
Sun. At that time the fragment configurations stay orbitally compact but encounters are beginning to
decrease as the orbital angles begin to spread. The peak value ${\bar p}_i\simeq 10^{-12}$ km$^{-2}$~yr$^{-1}$
is six orders of magnitude larger than the mean value over the main belt population. This value obviously
rapidly decreases in time, but at $500$~kyr, which is the current epoch for Datura family, it still attains
${\bar p}_i\simeq 1.38\times  10^{-15}$ km$^{-2}$~yr$^{-1}$, namely three orders of magnitude larger than the
mean value for the main belt. At face value, using this value alone, one would think that post family-formation
collisions cannot and should not be neglected.

However, the mean encounter velocities over the age of the Datura family are very small;
we find ${\bar v}\simeq 36$ m~s$^{-1}$, with the full range of $0.3$ to $500$ m~s$^{-1}$. As much as these values
are impressive,%
\footnote{We found it interesting to present some details of this numerical experiment, since we are not
 aware of a similar work previously published. It might be used as a template for studies of other
 very young families in the future.}
one may anticipate whether internal or external (main belt) impactors would be more important for the
Datura-family collisional evolution. The intrinsic collisional probability of Datura members between
each other is about three orders of magnitude larger than the probability
being hit by background main belt projectiles. However, the main belt impactor
population is about four orders of magnitude more numerous (Fig.~\ref{datura_sfd}).
Therefore, we expect that main belt projectiles will dominate collisional
evolution in the family.

Finally, it is useful to mention that catastrophic breakups are characterized by
the critical impact specific energy $Q^\star_D$, namely the energy per unit target mass delivered by the
projectile required for catastrophic disruption of the target (i.e., such that one-half the mass of the target
body escapes). Many studies have dealt with this important quantity \citep[see][for review]{betal2015},
but here we assume a simple relation (density $\rho$ also in cgs units)
\begin{equation}
  Q^\star_D = 9.0\times 10^7\,{\rm erg}\,{\rm g}^{-1} \left({D\over 2\,{\rm cm}}\right)^{-0.53}\!\!\! +
   0.5\,{\rm erg}\,{\rm {\rm cm}}^{-3}\rho\left({D\over 2\,{\rm cm}}\right)^{1.36}\!\!\!, \label{qs}
\end{equation}
whose constants have been adjusted provide a global stationary solution for the main belt asteroids
\cite[see also][for context]{bottke2020}. The critical size of a projectile able to catastrophically
disrupt a target of size $D$ scales as $\propto (Q^\star_D/{\bar v}^2)^{1/3}\,D$ and this minimizes  
the role of internal collisions  in  young families because impact velocities ${\bar v}$ are low.
.
\begin{figure}[t]
 \begin{center} 
 \includegraphics[width=0.47\textwidth]{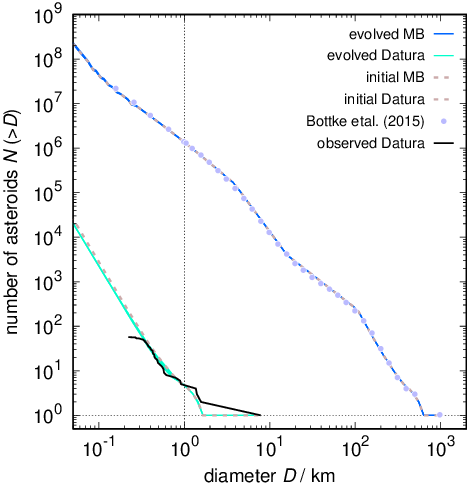}
 \end{center}
 \caption{Initial and final size distribution of the asteroid main belt (blue) and the synthetic Datura
  family (cyan) from the $500$~kyr lasting simulation of the {\tt Boulder} code. Intrinsic collisonal
  probabilities and impact velocities as described in the text. The observed Datura family population
  from Table~\ref{datura_members} shown by the black curve. No evolution is seen in both populations over
  the tested timescale and for the sizes displayed.}
 \label{datura_sfd}
 \end{figure}

We ran the {\tt Boulder} code for a $500$~kyr timespan and obtained results shown in Fig.~\ref{datura_sfd}.
No change in the family size distribution for $D>50$~m was  recorded, likely because the evolution timespan
was too short.  In families, the change in their size frequency distribution propagates from small to
larger sizes, and in $500$~kyr it only reaches the $\simeq 10$~cm range within the Datura family.
Similar results were obtained for Adelaide and Rampo families too.

Summing up the previous simulations, we conclude that collisional evolution over the timescales
corresponding to the ages of the very young families is not capable  of producing a transition to a shallower
segment of the family size distribution at about $300$-$400$~m. If true, the bias-corrected family population
from the current-date observations correspond also to the population of members created at the family
origin.
\smallskip

\noindent{\it Further results and future outlooks.-- }The estimated parameters of the magnitude
distributions obtained above may serve for additional consistency checks. For instance, assuming the
bias-corrected populations are representative of those generated right after disruption of the
parent body of the family at the observed sizes, we may use the estimated mass in small
members to determine further quantities. In the case of the Adelaide family we found 
$m_{\rm frag}/m_{\rm LF}\simeq 0.0085$. With that number, we may estimate (the minimum) size
of the crater on (525) Adelaide that has been formed. If we take crater depth to be $\simeq 1/10$-$1/5$
of its radius \citep[e.g.,][]{m1989}, and $D_{525} \simeq 9.4$~km the size of (525) Adelaide, a simple
calculation shows that a crater with $D_{\rm crat}\simeq 3.6$-$3.9$~km would have about the same
volume fraction in (525) Adelaide. This is still a reasonable number. Additionally, assuming
crater to projectile size ratio of $\simeq 10$-$20$ \citep[e.g.,][]{bottke2020}, we may estimate the
projectile size to about $d_{\rm proj}\simeq 180$-$390$~m. The number of 10~km size asteroids in the
inner main belt is $N_{10}\simeq 300$ \citep[e.g.,][]{masiero2011}, and the number of $180$-$390$~m objects
in the main belt $N_{\rm proj}\simeq (1$-$5)\times 10^7$ \citep[e.g.,][]{bottke2020}. Considering 
the mean intrinsic collision probability in the main belt $p_{\rm i}\simeq 2.8\times 10^{-18}$ km$^{-2}$
yr$^{-1}$, we may estimate the frequency of $180$-$390$~m projectiles impacting a $10$~km inner main belt
target to about $f\simeq p_{\rm i} R^2 N_{\rm proj} N_{10}\simeq (0.2$-$1)\times 10^{-6}$~yr$^{-1}$. This
results in a characteristic timescale of $\simeq 1$-$5$~Myr, which is well comparable with the
estimated age of the Adelaide family \citep[about $540$~kyr, e.g.,][]{adelaideAA2021}. While
highly simplified, this reasoning points to rough consistency between the Adelaide family origin and
the produced fragment population.
\smallskip

Very young asteroid families will certainly occupy interest of planetary scientists in the
forthcoming decade. While theoretical studies will continue, perhaps even more important input is expected
on the observational side. The planned powerful surveys, such as the Vera C. Rubin observatory
\citep[e.g.,][]{schwa2023}, promise to increase the known inventory of these clusters by an
order of magnitude, pushing the completeness near to the absolute magnitude $20$ (at least
for clusters in the inner main belt). Unlike the case of large and old asteroid families, the
identification of the very young families may be still a straightforward task (profiting from
the 5D arena of the osculating orbital elements and a possibility to recognize interlopers using
backward orbital propagation). The magnitude distribution of a complete population of members
may be set much more reliably, including the critical interval of $H$ in between $19$ and
$20$ magnitude.

\begin{acknowledgements}
 We thank the referee for very useful suggestions on the submitted version of the paper.
 We are grateful to the Catalina Sky Survey staff, Eric Christensen and Franck Shelly in particular,
 for providing us the CSS observations in between 2013 and 2022 in a user-friendly format and allowing
 us to use them for this research. This work was supported by the Czech Science Foundation (grant~21-11058S).
\end{acknowledgements}

\bibliographystyle{aa}

\begin{appendix}

\section{Members of the very young families}
Here we provide an information about membership in the very young asteroid families
studied in this paper. Our approach to obtain these results is based on two criteria.
First, we search for asteroids located in the vicinity of the largest member in
the 5D space of osculating orbital elements (disregarding longitude in orbit $\lambda$):
semimajor axis $a$, eccentricity $e$, inclination $I$, longitude of node $\Omega$,
and argument of perihelion $\omega$. Unlike in \citet{nv2006}, we do not use any specific
metric function, but simply select all asteroids with orbits in a certain box. In particular,
we let the orbital elements vary by the following limits: (i) semimajor axis by $\pm
0.03$~au, (ii)
eccentricity by $\pm 0.03$, (iii) inclination by $\pm 0.2^\circ$, and (iv) longitude
of node and argument of perihelion both by $\pm 30^\circ$. These values are larger than
the short-period oscillations of these elements due to planetary perturbations, and
conservative enough to sense the population even to the smallest currently detectable
sizes (note that small members might have been ejected with larger velocity than the
larger ones, constituting the family core). Given the large increase in number of discovered
asteroids, there is a small but nonzero chance that such a simple selection method may associate
background (unrelated) objects to the family even in the vast 5D space. For that reason,
we perform in the second step a convergence control. We numerically integrate orbits of
all identified asteroids backward in time for $2$~Myr. To keep things simple, we use
only nominal (best-fit) initial data at MJD epoch 60,000.0 and include only gravitational
perturbations from all planets (disregarding thermal accelerations).%
\footnote{We use a well-tested and publicly available integration package {\tt swift}
 (\url{http://www.boulder.swri.edu/~hal/swift.html})
 with a short timestep of $2$~d. We output the asteroid heliocentric state vectors every
 $5$~yr and monitor convergence of the secular angles $\Omega$ and $\varpi$ toward the
 reference values of the largest member in the family.}
The planetary configuration at the initial epoch is obtained from the JPL ephemerides
file DE 421. As a result, the purpose of this simulation is not to accurately determine
the age of the family, which is for most cases known from previous studies, but to eliminate
possible interloping objects. We found that the secular angles in the interloper cases show a
rapid divergence from the largest body in the family and may be easily identified.
Obviously, we eliminated these objects from our analysis of the size distribution of
the family members. There was only a limited amount of such objects found. The most
crowded situation occurred for Datura family, where we eliminated $80$ such objects, i.e., little
less than the family members (who are strongly clustered in the simple 5D box of orbital
elements that we considered for family-member search).
\onecolumn
\begin{longtable}{rlccccccc}
\caption{\label{datura_members}
 Datura family as of June 2023. Osculating heliocentric orbital elements at epoch
 MJD 60,000.0 from the MPC catalog: semimajor axis $a$, eccentricity $e$, inclination $I$,
 longitude of node $\Omega$, and argument of perihelion $\omega$. singleopposition orbits are
 listed at the end of the table. The third column gives the absolute magnitude $H$. The last
 column indicates, whether the asteroid has been detected by CSS during the phase~2
 operations (Y=yes). We note two very small, singleopposition asteroids 2016~PL51 and
 2022~RB57, very likely members of the Datura family too. However, their orbits, based on
 observations spanning short arcs (less than a week in the case of 2016~PL51), are still
 very uncertain. We include (429988) 2013~PZ36 residing on a rather chaotic orbit (most likely
 interacting with the exterior E3/10 mean motion resonance with the Earth), such that
 proving its membership to the Datura family would require an extensive work beyond the scope
 of this paper (see also Fig.~\ref{fd_1}). Luckily, the results discussed in Sec.~\ref{res} are not
 overly sensitive to the decision about Datura membership of this body.} \\
 \hline \hline
 \multicolumn{2}{c}{Asteroid} & \rule{0pt}{2ex} $H$ & $a$ & $e$ & $I$ & $\Omega$ & $\omega$ & CSS \\
	& & (mag) & (au) & & (deg) & (deg) & (deg) & \\ [1pt]
 \hline
 \endfirsthead
 \caption{continued.}\\
 \hline\hline
 \multicolumn{2}{c}{Asteroid} & \rule{0pt}{2ex} $H$ & $a$ & $e$ & $I$ & $\Omega$ & $\omega$ & CSS \\
	& & (mag) & (au) & & (deg) & (deg) & (deg)&  \\ [1pt]
 \hline
 \endhead
 \hline
 \endfoot
 \rule{0pt}{3ex}
   1270 & Datura     & 12.54 & 2.2344232 & 0.2080399 & 5.98629 & $\phantom{1}$97.77551 & 259.06896 & Y \\    
  60151 & 1999 UZ6   & 16.35 & 2.2347805 & 0.2078266 & 5.99466 & $\phantom{1}$96.68391 & 260.77419 & Y \\    
  89309 & 2001 VN36  & 16.48 & 2.2356444 & 0.2064328 & 6.01966 & $\phantom{1}$92.81960 & 267.16899 & Y \\    
  90265 & 2003 CL5   & 16.08 & 2.2347649 & 0.2074858 & 5.99650 & $\phantom{1}$95.55568 & 262.11246 & Y \\    
 203370 & 2001 WY35  & 17.33 & 2.2352682 & 0.2074548 & 5.98927 & $\phantom{1}$96.71788 & 260.76103 & Y \\    
 215619 & 2003 SQ168 & 17.24 & 2.2343739 & 0.2080248 & 5.98761 & $\phantom{1}$97.36200 & 259.62134 & Y \\    
 338309 & 2002 VR17  & 17.68 & 2.2345469 & 0.2077371 & 5.99159 & $\phantom{1}$96.68295 & 260.80932 & Y \\    
 429988 & 2013 PZ36  & 17.95 & 2.2306873 & 0.2109170 & 5.82996 & 102.85349 & 248.77348 & Y \\
 433382 & 2013 ST71  & 18.09 & 2.2343697 & 0.2079372 & 5.98521 & $\phantom{1}$97.92635 & 259.05669 & Y \\    
 452713 & 2005 YP136 & 18.46 & 2.2367817 & 0.2052885 & 6.04942 & $\phantom{1}$86.12897 & 276.54525 & Y \\    
 485010 & 2009 VS116 & 18.23 & 2.2361205 & 0.2051146 & 6.06774 & $\phantom{1}$85.69555 & 277.53365 & Y \\    
 553350 & 2011 KT10  & 18.15 & 2.2361705 & 0.2068784 & 6.02362 & $\phantom{1}$92.38721 & 267.95500 & Y \\    
 585600 & 2018 VR79  & 18.54 & 2.2350642 & 0.2074864 & 5.98893 & $\phantom{1}$97.32214 & 259.95573 & Y \\    
        & 2002 RH291 & 17.97 & 2.2349777 & 0.2076290 & 5.99691 & $\phantom{1}$95.64466 & 262.24539 & Y \\   
        & 2002 UU58  & 19.97 & 2.2347195 & 0.2074394 & 5.99799 & $\phantom{1}$96.59631 & 261.33505 & \\    
        & 2003 UD112 & 18.10 & 2.2347778 & 0.2073396 & 5.99912 & $\phantom{1}$95.41269 & 263.14733 & Y \\      
        & 2005 RK54  & 18.75 & 2.2348344 & 0.2066777 & 6.03197 & $\phantom{1}$92.65329 & 267.58009 & \\        
        & 2006 KA77  & 18.31 & 2.2343902 & 0.2083194 & 5.98156 & $\phantom{1}$99.49835 & 256.42782 & Y \\     
	& 2006 SY376 & 20.30 & 2.2329680 & 0.2096354 & 5.96933 & 107.19877 & 245.09390 & \\                  
        & 2006 SD382 & 18.91 & 2.2361239 & 0.2055712 & 6.05939 & $\phantom{1}$85.86581 & 277.02849 & Y \\    
        & 2006 WV222 & 18.80 & 2.2350979 & 0.2080650 & 5.98696 & $\phantom{1}$97.78267 & 259.39354 & Y \\      
        & 2007 RM332 & 18.47 & 2.2351994 & 0.2077589 & 5.98292 & $\phantom{1}$98.46057 & 258.26665 & Y \\     
        & 2008 YV51  & 18.60 & 2.2353599 & 0.2075360 & 5.98733 & $\phantom{1}$97.48832 & 259.71301 & Y \\     
        & 2010 VN260 & 19.20 & 2.2349143 & 0.2073926 & 5.99770 & $\phantom{1}$96.14005 & 262.02148 & Y \\     
        & 2010 VU261 & 19.10 & 2.2346754 & 0.2076944 & 5.99159 & $\phantom{1}$96.65770 & 260.93281 & Y \\    
        & 2010 VB265 & 19.10 & 2.2348391 & 0.2075374 & 5.99878 & $\phantom{1}$96.29429 & 261.51547 & Y \\     
        & 2012 VN143 & 19.32 & 2.2346189 & 0.2074989 & 5.99550 & $\phantom{1}$96.99693 & 260.66846 & \\      
        & 2014 NZ88  & 18.80 & 2.2353179 & 0.2072581 & 5.98828 & $\phantom{1}$96.96231 & 260.64555 & Y \\    
	& 2014 OY85  & 19.50 & 2.2343694 & 0.2085063 & 5.97725 & 100.94642 & 253.92579 & \\                   
	& 2014 OA86  & 18.87 & 2.2349119 & 0.2078508 & 5.97656 & 100.39381 & 255.53971 & Y \\                
        & 2014 OE206 & 19.26 & 2.2354740 & 0.2070515 & 5.99678 & $\phantom{1}$96.10416 & 262.05661 & Y \\     
        & 2014 OR378 & 18.77 & 2.2352076 & 0.2075890 & 5.98585 & $\phantom{1}$97.77828 & 259.17541 & Y \\    
        & 2014 WL96  & 19.30 & 2.2368677 & 0.2067078 & 6.00707 & $\phantom{1}$93.95385 & 265.59361 & \\      
        & 2014 WT96  & 18.93 & 2.2355619 & 0.2072284 & 5.99606 & $\phantom{1}$95.93259 & 262.19506 & \\      
        & 2015 DY94  & 18.20 & 2.2346085 & 0.2075363 & 5.99356 & $\phantom{1}$96.48915 & 261.34069 & Y \\    
        & 2015 PD191 & 20.00 & 2.2360619 & 0.2070593 & 6.02521 & $\phantom{1}$93.22656 & 266.76523 & \\ 
        & 2015 PQ47  & 19.17 & 2.2343287 & 0.2077176 & 5.99164 & $\phantom{1}$97.27727 & 259.91321 & \\      
        & 2015 PH144 & 19.56 & 2.2337492 & 0.2084264 & 5.98369 & $\phantom{1}$98.81562 & 257.09515 & \\      
        & 2015 PR301 & 18.87 & 2.2343558 & 0.2078212 & 5.98601 & $\phantom{1}$98.26718 & 258.53367 & \\      
        & 2015 QW31  & 19.00 & 2.2342126 & 0.2078737 & 5.98459 & $\phantom{1}$98.21808 & 258.43923 & \\      
        & 2015 SS31  & 18.84 & 2.2351942 & 0.2063313 & 6.00527 & $\phantom{1}$91.16595 & 268.71019 & Y \\      
        & 2015 TL455 & 18.59 & 2.2345918 & 0.2076533 & 5.99230 & $\phantom{1}$97.58170 & 259.65709 & Y \\     
        & 2015 WQ25  & 18.70 & 2.2347622 & 0.2074942 & 5.99209 & $\phantom{1}$96.92848 & 260.47264 & Y \\    
        & 2015 XK88  & 18.63 & 2.2341292 & 0.2082101 & 5.98363 & $\phantom{1}$99.13007 & 256.75539 & Y \\    
        & 2015 XX321 & 19.13 & 2.2349203 & 0.2073029 & 5.99831 & $\phantom{1}$96.45517 & 261.40030 & Y \\    
        & 2015 XQ432 & 18.80 & 2.2347230 & 0.2073054 & 5.99311 & $\phantom{1}$96.44434 & 261.32662 & Y \\    
        & 2015 XK452 & 18.98 & 2.2348068 & 0.2074621 & 5.99415 & $\phantom{1}$97.07086 & 260.30732 & Y \\    
        & 2016 TW15  & 18.70 & 2.2346167 & 0.2077361 & 5.99193 & $\phantom{1}$97.14749 & 260.35782 & Y \\    
        & 2016 TR115 & 18.70 & 2.2348542 & 0.2078892 & 5.98988 & $\phantom{1}$97.68880 & 259.59729 & Y \\     
        & 2017 QX88  & 18.88 & 2.2370037 & 0.2053516 & 6.06398 & $\phantom{1}$85.78189 & 277.15110 & Y \\     
        & 2017 SU3   & 18.91 & 2.2352968 & 0.2074508 & 5.98488 & $\phantom{1}$97.47714 & 259.99874 & Y \\     
        & 2017 SV143 & 18.80 & 2.2354487 & 0.2049571 & 6.06182 & $\phantom{1}$86.32039 & 276.61565 & Y \\    
	    & 2017 UW137 & 19.75 & 2.2343672 & 0.2084085 & 5.97502 & 101.13381 & 254.12998 & Y \\                 
        & 2017 UU155 & 19.62 & 2.2365893 & 0.2069006 & 6.00884 & $\phantom{1}$93.82864 & 265.81621 & Y \\     
        & 2017 VP37  & 19.40 & 2.2352166 & 0.2075381 & 5.99312 & $\phantom{1}$97.08958 & 260.46478 & Y \\      
        & 2017 WC50  & 19.53 & 2.2352780 & 0.2073674 & 5.99337 & $\phantom{1}$96.33838 & 261.68738 & \\         
        & 2018 TM7   & 18.71 & 2.2347907 & 0.2075211 & 5.98873 & $\phantom{1}$97.80903 & 259.29224 & Y \\    
        & 2018 UN34  & 19.25 & 2.2352820 & 0.2068990 & 5.99899 & $\phantom{1}$96.01196 & 262.59590 & Y \\    
        & 2018 UL40  & 19.10 & 2.2353244 & 0.2069677 & 6.00250 & $\phantom{1}$96.17580 & 262.28729 & Y \\    
        & 2019 QA14  & 18.60 & 2.2351143 & 0.2078881 & 5.99236 & $\phantom{1}$97.28433 & 260.08707 & Y \\    
        & 2019 SE28  & 19.16 & 2.2342820 & 0.2087329 & 5.98412 & $\phantom{1}$99.30327 & 256.39995 & Y \\    
        & 2019 XJ15  & 19.12 & 2.2350562 & 0.2073834 & 6.00613 & $\phantom{1}$95.41326 & 263.12590 & Y \\    
        & 2020 OS89  & 19.54 & 2.2348379 & 0.2075650 & 6.00508 & $\phantom{1}$93.45550 & 264.93955 & \\            
        & 2020 PM28  & 19.24 & 2.2352071 & 0.2077160 & 5.99066 & $\phantom{1}$97.81753 & 259.47920 & Y \\            
        & 2021 RB114 & 18.60 & 2.2352733 & 0.2073676 & 5.99044 & $\phantom{1}$96.69883 & 260.81727 & Y \\            
        & 2022 QC148 & 19.50 & 2.2343589 & 0.2078442 & 5.99308 & $\phantom{1}$97.08424 & 260.07491 & \\            
        & 2022 SV168 & 19.59 & 2.2348466 & 0.2051558 & 6.06436 & $\phantom{1}$86.74092 & 275.74315 & \\ [6pt]           
             \multicolumn{9}{c}{-- Singleopposition members --} \\                    
             \rule{0pt}{3ex}                                                           
        & 2014 WG250 & 18.95 & 2.2352239 & 0.2075190 & 5.98785 & $\phantom{1}$97.36655 & 259.84769 & \\            
        & 2014 WM249 & 19.19 & 2.2339095 & 0.2075368 & 5.98235 & $\phantom{1}$97.79023 & 259.01686 & \\             
        & 2015 TU306 & 19.60 & 2.2349386 & 0.2067806 & 6.01564 & $\phantom{1}$95.68816 & 263.34206 & \\            
        & 2016 PY22  & 19.44 & 2.2360027 & 0.2069173 & 6.02658 & $\phantom{1}$92.76382 & 267.85893 & \\
        & 2017 OS162 & 19.49 & 2.2360582 & 0.2070355 & 5.98009 & $\phantom{1}$90.91999 & 267.41699 & \\ 
        & 2017 OU162 & 19.69 & 2.2353099 & 0.2074271 & 5.98822 & $\phantom{1}$97.34285 & 260.05416 & \\              
        & 2017 SG152 & 19.00 & 2.2357888 & 0.2064416 & 6.01963 & $\phantom{1}$93.43504 & 266.60818 & Y \\              
        & 2017 SV193 & 19.60 & 2.2352198 & 0.2074399 & 5.99136 & $\phantom{1}$96.97981 & 260.72387 & Y \\             
        & 2017 SC233 & 19.20 & 2.2350514 & 0.2074005 & 6.00705 & $\phantom{1}$93.16523 & 265.47990 & \\             
        & 2017 SS269 & 19.90 & 2.2360231 & 0.2070277 & 6.00573 & $\phantom{1}$95.79973 & 263.47065 & \\             
        & 2019 TD28  & 19.60 & 2.2368243 & 0.2075650 & 6.01491 & $\phantom{1}$94.64557 & 265.09847 & Y \\             
        & 2020 QM36  & 19.00 & 2.2349968 & 0.2076282 & 5.98836 & $\phantom{1}$97.10501 & 260.48247 & Y \\             
        & 2020 RR103 & 19.80 & 2.2361572 & 0.2077930 & 5.98241 & $\phantom{1}$90.88493 & 267.16232 & Y \\      
        & 2020 UV37  & 19.30 & 2.2354422 & 0.2074241 & 5.97418 & $\phantom{1}$90.97581 & 267.30544 & Y \\      
        & 2021 NF47  & 19.49 & 2.2350049 & 0.2075666 & 5.98830 & $\phantom{1}$97.42570 & 259.67732 & Y\\       
        & 2021 NK57  & 19.14 & 2.2353946 & 0.2069712 & 5.99919 & $\phantom{1}$95.52787 & 262.99252 & Y \\      
        & 2021 PX107 & 19.22 & 2.2352389 & 0.2073418 & 5.98930 & $\phantom{1}$97.13408 & 260.30247 & Y \\      
        & 2021 QZ40  & 19.75 & 2.2350184 & 0.2074293 & 5.99078 & $\phantom{1}$96.79357 & 260.66997 & \\
        & 2021 RE149 & 19.00 & 2.2349079 & 0.2077411 & 5.98546 & $\phantom{1}$98.19857 & 258.44555 & Y \\    
        & 2021 VU20  & 20.32 & 2.2355854 & 0.2063734 & 6.00005 & $\phantom{1}$95.48181 & 263.86595 & \\             
        & 2022 PN15  & 19.82 & 2.2352065 & 0.2070323 & 6.02051 & $\phantom{1}$94.91401 & 264.61153 & \\             
        & 2022 QK69  & 19.85 & 2.2361785 & 0.2073031 & 5.99880 & $\phantom{1}$95.16386 & 263.81182 & \\             
        & 2022 QT171 & 19.62 & 2.2363260 & 0.2071205 & 6.02165 & $\phantom{1}$94.40238 & 265.28039 & \\             
	& 2022 SO76  & 19.43 & 2.2339303 & 0.2084255 & 5.98316 & $\phantom{1}$98.95475 & 256.96663 & \\
	& 2022 TV22  & 20.14 & 2.2362992 & 0.2055041 & 6.06042 & $\phantom{1}$85.66342 & 276.96670 & \\ [2pt] 
 \hline
\end{longtable}

\onecolumn
\begin{longtable}{rlccccccc}
\caption{\label{adelaide_members}
 Adelaide family as of June  2023. Osculating heliocentric orbital elements at epoch
 MJD 60,000.0 from the MPC catalog: semimajor axis $a$, eccentricity $e$, inclination $I$,
 longitude of node $\Omega$, and argument of perihelion $\omega$. singleopposition orbits are
 listed at the end of the table. The third column gives the absolute magnitude $H$. The last
 column indicates, whether the asteroid has been detected by CSS during the phase~2
 operations (Y=yes). We note asteroid (159941) 2005~WV178 in the near vicinity of the Adelaide
 family, which we discard from the membership due to a dubious convergence to (525) Adelaide
 in the past Myr.} \\
 \hline \hline
 \multicolumn{2}{c}{Asteroid} & \rule{0pt}{2ex} $H$ & $a$ & $e$ & $I$ & $\Omega$ & $\omega$ & CSS \\
	& & (mag) & (au) & & (deg) & (deg) & (deg) & \\ [1pt]
 \hline
 \endfirsthead
 \caption{continued.}\\
 \hline\hline
 \multicolumn{2}{c}{Asteroid} & \rule{0pt}{2ex} $H$ & $a$ & $e$ & $I$ & $\Omega$ & $\omega$ & CSS \\
	& & (mag) & (au) & & (deg) & (deg) & (deg) & \\ [1pt]
 \hline
 \endhead
 \hline
 \endfoot
 \rule{0pt}{3ex}
    525 & Adelaide   & 12.17 & 2.2459455 & 0.1020388 & 5.99835 & 203.35936 & 263.96182 & Y \\     
 422494 & 2014 SV342 & 18.37 & 2.2457873 & 0.1036634 & 6.01316 & 201.90211 & 262.11221 & Y \\     
 452322 & 2000 GG121 & 18.43 & 2.2459962 & 0.0990051 & 6.05840 & 197.12857 & 277.42396 & Y \\     
 463394 & 2013 GV28  & 18.56 & 2.2452544 & 0.1014759 & 6.00185 & 203.34261 & 265.44077 & Y \\     
 475474 & 2006 SZ152 & 18.58 & 2.2446196 & 0.1025679 & 5.98782 & 204.84491 & 263.14855 & Y \\     
 486081 & 2012 UX41  & 18.54 & 2.2458709 & 0.1018798 & 6.01195 & 200.73242 & 265.92554 & Y \\     
 504375 & 2007 VV73  & 18.76 & 2.2455021 & 0.1034966 & 6.01373 & 200.64920 & 264.76138 & Y \\     
 517580 & 2014 UZ170 & 18.68 & 2.2457137 & 0.1015922 & 6.00769 & 200.32247 & 269.68162 & Y \\     
 534611 & 2014 UC204 & 18.18 & 2.2458760 & 0.1010302 & 6.02623 & 199.84308 & 272.15198 & Y \\     
 545614 & 2011 SA45  & 18.41 & 2.2457694 & 0.1022064 & 6.00750 & 201.81455 & 264.31208 & Y \\     
 552867 & 2010 UF125 & 18.78 & 2.2449504 & 0.1025609 & 6.01004 & 202.13835 & 266.85951 & Y \\     
 555571 & 2014 AD31  & 18.51 & 2.2457958 & 0.1009470 & 6.02260 & 200.55360 & 270.49477 & Y \\     
 569552 & 2005 UK370 & 19.01 & 2.2455102 & 0.1014688 & 6.02424 & 200.81248 & 268.01004 & Y \\     
 572830 & 2008 US17  & 18.57 & 2.2455380 & 0.1000395 & 6.04296 & 198.32560 & 273.73347 & Y \\     
 572868 & 2008 UR182 & 18.40 & 2.2458141 & 0.1034843 & 6.02337 & 200.30854 & 263.50713 & Y \\     
 578969 & 2014 JA2   & 18.26 & 2.2457211 & 0.1010352 & 6.03014 & 199.49735 & 269.87273 & Y \\     
 593790 & 2015 XZ90  & 18.70 & 2.2449726 & 0.1014565 & 5.99848 & 203.78817 & 264.25467 & Y \\     
 616487 & 2005 VP83  & 18.45 & 2.2456512 & 0.1016943 & 6.02430 & 200.22763 & 267.30317 & Y \\     
        & 2004 HU76  & 19.02 & 2.2462293 & 0.1005166 & 6.00980 & 202.63100 & 267.20087 & Y \\     
        & 2004 HJ85  & 18.97 & 2.2461326 & 0.1032087 & 6.03593 & 200.54895 & 263.66537 & Y \\     
        & 2005 UF193 & 18.77 & 2.2454086 & 0.1038225 & 6.01911 & 200.93875 & 261.58535 & Y \\     
        & 2006 SK449 & 18.40 & 2.2456295 & 0.1013371 & 6.02272 & 200.42520 & 271.19772 & Y \\     
        & 2007 TA504 & 18.90 & 2.2455486 & 0.1024368 & 6.02680 & 200.28214 & 268.89512 & Y \\     
        & 2007 VT345 & 18.58 & 2.2449214 & 0.1035204 & 5.99871 & 203.77264 & 262.38520 & Y \\     
        & 2008 ET179 & 18.60 & 2.2450348 & 0.1025046 & 6.00338 & 202.63850 & 266.26648 & Y \\     
        & 2008 UR414 & 18.63 & 2.2460790 & 0.0989092 & 6.02968 & 201.69890 & 272.77329 & Y \\     
        & 2009 WJ157 & 18.77 & 2.2464285 & 0.1016017 & 6.01107 & 202.07146 & 265.29561 & Y \\     
        & 2010 VC228 & 18.33 & 2.2450406 & 0.1014959 & 6.03451 & 198.81306 & 273.39747 & Y \\     
        & 2010 VF260 & 18.60 & 2.2450148 & 0.1034408 & 6.04766 & 197.20466 & 270.60213 & Y \\     
        & 2010 XB115 & 18.81 & 2.2448215 & 0.1031887 & 6.00795 & 201.45005 & 265.44057 & Y \\     
        & 2012 TM342 & 19.30 & 2.2451705 & 0.1015429 & 6.00110 & 202.85893 & 265.02778 & \\     
        & 2013 CH251 & 19.43 & 2.2457832 & 0.1021203 & 6.00697 & 202.66417 & 264.04047 & Y \\   
        & 2013 GR162 & 18.80 & 2.2457624 & 0.1026941 & 6.00444 & 202.30810 & 263.28291 & Y \\   
        & 2013 HB97  & 19.80 & 2.2463044 & 0.1003166 & 6.02739 & 200.10111 & 272.11349 & \\     
        & 2013 TY219 & 18.96 & 2.2442386 & 0.1035677 & 5.99071 & 204.30989 & 262.02986 & Y \\   
        & 2013 TR236 & 19.54 & 2.2454223 & 0.1028989 & 6.01398 & 202.86209 & 264.25702 & Y \\   
        & 2014 EQ81  & 19.10 & 2.2469478 & 0.0999236 & 6.05462 & 197.32994 & 275.67967 & Y \\   
        & 2014 EU96  & 19.20 & 2.2464881 & 0.1040363 & 6.01320 & 200.66878 & 262.24741 & Y \\       
        & 2014 EM164 & 18.98 & 2.2459219 & 0.1031170 & 5.99009 & 204.47170 & 260.80897 & Y \\       
        & 2014 JY105 & 19.10 & 2.2459512 & 0.0987461 & 6.03208 & 201.11692 & 273.10746 & Y \\   
        & 2014 WM167 & 18.95 & 2.2458331 & 0.1018454 & 6.00962 & 202.17318 & 267.55573 & Y \\   
        & 2015 BE285 & 19.05 & 2.2449165 & 0.1039514 & 5.99999 & 203.11323 & 261.33797 & Y \\   
        & 2015 HU72  & 18.95 & 2.2452445 & 0.1031607 & 6.00944 & 202.19152 & 264.15943 & Y \\   
        & 2015 RM186 & 18.73 & 2.2460358 & 0.0984971 & 6.06417 & 196.43719 & 279.45108 & Y \\   
        & 2015 TD44  & 19.26 & 2.2444571 & 0.1025243 & 5.98632 & 204.98118 & 260.23068 & Y \\   
        & 2015 UR18  & 19.30 & 2.2454281 & 0.1021862 & 6.03178 & 200.65949 & 265.28313 & Y \\   
        & 2015 XC92  & 19.07 & 2.2460168 & 0.1009429 & 6.04263 & 200.06013 & 269.07119 & \\     
        & 2016 AH353 & 19.77 & 2.2460923 & 0.1013139 & 6.01534 & 202.02955 & 267.90961 & \\           
        & 2016 AL322 & 18.90 & 2.2456315 & 0.1026982 & 6.01532 & 202.92759 & 262.78436 & Y \\   
        & 2016 CP95  & 19.33 & 2.2463603 & 0.1003278 & 6.02127 & 200.61828 & 271.25649 & \\       
        & 2016 CX104 & 19.24 & 2.2448809 & 0.1035784 & 5.98559 & 205.29467 & 258.78148 & Y \\     
        & 2016 EX318 & 19.50 & 2.2449848 & 0.1030240 & 5.98796 & 204.87999 & 260.79999 & Y \\    
        & 2016 FR33  & 19.10 & 2.2458347 & 0.1015699 & 6.05834 & 196.65477 & 274.01002 & Y \\    
        & 2016 FA34  & 18.86 & 2.2455945 & 0.1024636 & 6.01441 & 201.96964 & 265.06660 & Y \\    
        & 2016 GO11  & 18.72 & 2.2450053 & 0.1037665 & 5.98548 & 205.07788 & 259.12530 & Y \\    
        & 2016 QE71  & 18.40 & 2.2454966 & 0.1029550 & 6.01495 & 201.16100 & 266.00988 & Y \\    
        & 2016 TN41  & 18.90 & 2.2446126 & 0.1043580 & 5.98344 & 205.11808 & 258.91716 & \\      
        & 2016 UO110 & 19.06 & 2.2454466 & 0.1026978 & 5.99586 & 203.55032 & 263.93640 & Y \\
        & 2017 AU38  & 18.78 & 2.2458978 & 0.1013304 & 6.00496 & 202.56750 & 266.13825 & Y \\    
        & 2017 HL72  & 19.28 & 2.2459566 & 0.0998460 & 6.02303 & 200.55185 & 271.45608 & Y \\   
        & 2017 RS100 & 19.33 & 2.2449371 & 0.1039013 & 6.00073 & 204.79324 & 259.00262 & Y \\   
        & 2017 TG26  & 18.87 & 2.2448290 & 0.1044112 & 5.98989 & 204.40015 & 259.95117 & Y \\   
        & 2017 UF65  & 19.32 & 2.2446513 & 0.1035296 & 6.00813 & 202.33968 & 263.80059 & Y \\   
        & 2017 WP50  & 19.10 & 2.2444171 & 0.1030306 & 6.00350 & 203.01479 & 264.40475 & \\     
        & 2019 BT11  & 19.11 & 2.2453842 & 0.1009102 & 6.02325 & 200.50607 & 270.84753 & Y \\   
        & 2019 TC62  & 19.20 & 2.2471638 & 0.0996640 & 6.05763 & 196.55998 & 277.07321 & Y \\  
        & 2019 YE29  & 19.77 & 2.2457962 & 0.1006684 & 6.02946 & 200.43589 & 269.04712 & Y \\  
        & 2019 YU35  & 19.86 & 2.2460175 & 0.1008068 & 6.02944 & 202.15923 & 266.45230 & \\           
        & 2020 ML45  & 19.00 & 2.2459836 & 0.1010309 & 6.02237 & 201.82072 & 267.71323 & \\    
        & 2020 PM79  & 19.50 & 2.2452151 & 0.1027546 & 6.00307 & 204.32861 & 262.16107 & \\    
        & 2022 BM6   & 19.30 & 2.2454318 & 0.1021906 & 6.00621 & 202.54005 & 266.53136 & Y \\  
        & 2022 CU16  & 18.98 & 2.2449320 & 0.1035621 & 6.01424 & 202.07267 & 263.45712 & Y \\          
	& 2022 TC6   & 18.99 & 2.2453803 & 0.1013650 & 6.00377 & 203.09324 & 265.02593 & \\ [6pt] 
             \multicolumn{8}{c}{-- Singleopposition members --} \\                    
             \rule{0pt}{3ex}                                                           
        & 2022 BM50  & 19.97 & 2.2451736 & 0.1020011 & 6.00018 & 201.42183 & 268.13980 & Y \\           
        & 2023 AH4   & 20.32 & 2.2458161 & 0.1016177 & 6.00306 & 203.14933 & 265.55282 & \\           
        & 2023 BX6   & 20.23 & 2.2454761 & 0.1028669 & 5.99652 & 203.75734 & 262.07433 & \\           
        & 2023 BZ6   & 20.11 & 2.2461186 & 0.1005071 & 6.01469 & 201.61036 & 269.59264 & \\           
        & 2023 BP9   & 20.53 & 2.2493354 & 0.1021673 & 6.02529 & 199.84371 & 272.57419 & \\           
        & 2023 BS11  & 20.57 & 2.2470465 & 0.1005654 & 6.01926 & 200.54636 & 271.10920 & \\ [2pt]
 \hline
\end{longtable}

\onecolumn
\begin{longtable}{rlccccccc}
\caption{\label{hobson_members}
 Hobson family as of June  2023. Osculating heliocentric orbital elements at epoch
 MJD 60,200.0 from the MPC catalog: semimajor axis $a$, eccentricity $e$, inclination $I$,
 longitude of node $\Omega$, and argument of perihelion $\omega$. singleopposition orbits are
 listed at the end of the table. The third column gives the absolute magnitude $H$. The last
 column indicates, whether the asteroid has been detected by CSS during the phase~2
 operations (Y=yes). We note a very small, singleopposition asteroids 2019~NF93, 2021~JQ73, 
 2023~JD27 and 2023~NV2
 very likely members of the Hobson family too. However, their orbits, especially for 2019~NF93
 based on observations spanning less than a week, are still very uncertain.} \\
 \hline \hline
 \multicolumn{2}{c}{Asteroid} & \rule{0pt}{2ex} $H$ & $a$ & $e$ & $I$ & $\Omega$ & $\omega$ & CSS \\
	& & (mag) & (au) & & (deg) & (deg) & (deg) & \\ [1pt]
 \hline
 \endfirsthead
 \caption{continued.}\\
 \hline\hline
 \multicolumn{2}{c}{Asteroid} & \rule{0pt}{2ex} $H$ & $a$ & $e$ & $I$ & $\Omega$ & $\omega$ & CSS \\
	& & (mag) & (au) & & (deg) & (deg) & (deg) & \\ [1pt]
 \hline
 \endhead
 \hline
 \endfoot
 \rule{0pt}{3ex}
  18777 & Hobson     & 15.12 & 2.5633566 & 0.1833929 & 4.32167 & 105.43986 & 180.62462 & Y \\                               
  57738 & 2001 UZ160 & 15.27 & 2.5643655 & 0.1804590 & 4.31695 & 104.86638 & 181.39287 & Y \\                               
 363118 & 2001 NH14  & 17.34 & 2.5640791 & 0.1802580 & 4.31088 & 105.05377 & 181.32416 & Y \\                               
 381414 & 2008 JK37  & 17.70 & 2.5644935 & 0.1801728 & 4.32092 & 104.22939 & 181.70561 & Y \\                               
 436620 & 2011 LF12  & 17.35 & 2.5623364 & 0.1840800 & 4.32629 & 104.88486 & 180.38868 & Y \\                               
 450571 & 2006 JH35  & 17.64 & 2.5622205 & 0.1830388 & 4.31799 & 105.19338 & 180.53430 & Y \\                               
 465404 & 2008 HQ46  & 17.67 & 2.5640097 & 0.1819144 & 4.31527 & 105.23235 & 182.39770 & Y \\                               
 520394 & 2014 JJ10  & 18.15 & 2.5636411 & 0.1819023 & 4.31680 & 105.02491 & 180.59212 & Y \\                               
 537249 & 2015 HM190 & 17.60 & 2.5618837 & 0.1853909 & 4.32936 & 105.06573 & 181.29952 & Y \\                               
 548822 & 2010 VG231 & 18.08 & 2.5645095 & 0.1785389 & 4.30907 & 104.44899 & 180.18738 & Y \\                               
 557505 & 2014 UB262 & 18.33 & 2.5644669 & 0.1812880 & 4.30940 & 105.44687 & 181.70689 & Y \\                               
        & 2007 EH116 & 17.60 & 2.5632497 & 0.1837133 & 4.33016 & 104.12508 & 181.78781 & Y \\                               
        & 2007 HC54  & 17.10 & 2.5630025 & 0.1852376 & 4.33060 & 103.90562 & 183.45131 & Y \\                               
        & 2008 WV149 & 18.25 & 2.5616506 & 0.1860344 & 4.32935 & 105.32362 & 181.64195 & Y \\                               
        & 2009 SY179 & 18.10 & 2.5638560 & 0.1808079 & 4.31363 & 105.25771 & 181.95983 & \\                                 
        & 2010 GN203 & 18.19 & 2.5616128 & 0.1827647 & 4.31768 & 105.51576 & 178.85422 & Y \\                               
        & 2011 SU302 & 18.40 & 2.5613443 & 0.1843413 & 4.32594 & 105.00690 & 180.83267 & Y \\                               
        & 2012 JM71  & 18.34 & 2.5643906 & 0.1803477 & 4.31944 & 104.45794 & 181.49202 & Y \\                               
        & 2012 LN31  & 18.15 & 2.5645366 & 0.1805664 & 4.32084 & 104.18716 & 181.60272 & Y \\                               
        & 2013 JG48  & 18.42 & 2.5640127 & 0.1804333 & 4.31002 & 105.27028 & 181.30120 & \\                                 
        & 2013 MW20  & 18.10 & 2.5640282 & 0.1789098 & 4.30379 & 105.82861 & 179.66519 & Y \\                               
        & 2013 NA73  & 17.90 & 2.5645573 & 0.1778761 & 4.31150 & 104.20836 & 180.13663 & Y \\                               
        & 2014 HH103 & 17.96 & 2.5628983 & 0.1818845 & 4.31303 & 105.17981 & 179.85773 & Y \\                               
        & 2014 KY102 & 18.08 & 2.5643135 & 0.1802754 & 4.30634 & 105.52954 & 178.70247 & \\                                 
        & 2014 NN71  & 18.22 & 2.5656862 & 0.1796550 & 4.31314 & 104.34699 & 180.71664 & \\                                 
        & 2014 OG277 & 18.40 & 2.5655560 & 0.1824004 & 4.30982 & 105.58409 & 182.32744 & \\                                 
        & 2014 OJ66  & 18.94 & 2.5662203 & 0.1795922 & 4.30905 & 105.02988 & 179.79448 & \\                                 
        & 2014 PJ87  & 18.30 & 2.5657155 & 0.1814269 & 4.31509 & 105.41914 & 181.45122 & \\                                 
        & 2014 QL520 & 18.41 & 2.5655987 & 0.1802430 & 4.30661 & 105.04757 & 180.77744 & \\                                 
        & 2014 QQ580 & 18.83 & 2.5657832 & 0.1791684 & 4.31163 & 104.58123 & 180.17054 & \\                                 
        & 2015 FV225 & 17.60 & 2.5626639 & 0.1856256 & 4.32625 & 105.38305 & 182.32173 & Y \\                               
        & 2015 HV138 & 18.70 & 2.5624216 & 0.1841988 & 4.32970 & 104.43650 & 181.05676 & \\                                 
        & 2015 KA91  & 17.90 & 2.5623849 & 0.1834933 & 4.32926 & 104.19929 & 180.34602 & Y \\                               
        & 2015 KM237 & 19.48 & 2.5623846 & 0.1836411 & 4.33279 & 103.88926 & 180.72858 & \\                                 
        & 2015 OP104 & 18.00 & 2.5614614 & 0.1838197 & 4.32310 & 104.57302 & 180.62155 & Y \\                               
        & 2015 PM156 & 18.40 & 2.5619337 & 0.1823281 & 4.32222 & 104.32621 & 179.28149 & \\                                 
        & 2015 PA184 & 19.20 & 2.5607541 & 0.1873221 & 4.32194 & 105.96144 & 182.84203 & \\                                 
        & 2015 XL282 & 17.79 & 2.5657472 & 0.1813020 & 4.31120 & 105.23265 & 181.11965 & Y \\                               
        & 2016 GY256 & 18.24 & 2.5636554 & 0.1832090 & 4.32326 & 105.46202 & 182.61057 & Y \\                               
        & 2016 GW276 & 18.48 & 2.5640705 & 0.1811033 & 4.31689 & 105.07693 & 181.94298 & Y \\                               
        & 2016 GZ310 & 18.51 & 2.5642163 & 0.1812362 & 4.32011 & 104.87910 & 181.95655 & Y \\                               
        & 2017 PA68  & 18.20 & 2.5644305 & 0.1794933 & 4.31098 & 104.94606 & 180.90470 & \\                                 
        & 2017 PK70  & 18.80 & 2.5631898 & 0.1834456 & 4.31379 & 105.99206 & 183.69467 & \\                                 
        & 2017 SM25  & 18.75 & 2.5641029 & 0.1805951 & 4.31554 & 104.93415 & 181.75945 & \\                                 
        & 2017 SQ83  & 18.33 & 2.5641221 & 0.1801750 & 4.31364 & 105.58058 & 180.54802 & Y \\                               
        & 2017 WO47  & 18.12 & 2.5639686 & 0.1820342 & 4.32084 & 104.97053 & 181.94034 & Y \\                               
        & 2018 NQ48  & 18.79 & 2.5649606 & 0.1809579 & 4.31287 & 105.27527 & 181.36440 & \\                                 
        & 2019 NP44  & 18.90 & 2.5614875 & 0.1835832 & 4.32402 & 104.94548 & 180.28464 & Y \\                               
        & 2019 NB193 & 19.09 & 2.5616801 & 0.1826859 & 4.32590 & 104.31100 & 179.86577 &   \\                               
        & 2019 PS30  & 18.50 & 2.5613773 & 0.1841029 & 4.32102 & 105.34292 & 180.47624 & Y \\                               
	& 2020 HQ57  & 18.50 & 2.5648348 & 0.1794839 & 4.31150 & 104.69543 & 180.06419 & Y \\                                 
        & 2020 KP36  & 19.11 & 2.5648730 & 0.1791752 & 4.31520 & 104.47316 & 179.58657 & \\
        & 2021 MO5   & 19.07 & 2.5636887 & 0.1815050 & 4.31175 & 105.50891 & 182.38281 & \\      
        & 2023 JA22  & 18.24 & 2.5617572 & 0.1834690 & 4.32797 & 104.29884 & 180.14545 & \\ [6pt]  
             \multicolumn{8}{c}{-- Singleopposition members --} \\
             \rule{0pt}{3ex}                                       
        & 2014 JH120 & 18.70 & 2.5642333 & 0.1818812 & 4.31659 & 105.15400 & 180.80286 & \\             
        & 2017 NY29  & 18.95 & 2.5644299 & 0.1787774 & 4.31392 & 104.20540 & 181.00297 & \\             
        & 2019 GR115 & 18.80 & 2.5620456 & 0.1857152 & 4.32843 & 105.21081 & 181.52184 & \\             
        & 2020 JM31  & 18.50 & 2.5636410 & 0.1836011 & 4.32256 & 105.32188 & 182.52160 & \\             
        & 2020 OY50  & 18.60 & 2.5626675 & 0.1856940 & 4.32491 & 105.47980 & 182.45254 & \\             
        & 2023 JZ8   & 18.67 & 2.5611412 & 0.1864161 & 4.32528 & 105.35980 & 181.88047 & \\ [2pt]       
 \hline
\end{longtable}

\onecolumn
\begin{longtable}{rlccccccc}
\caption{\label{rampo_members}
 Rampo family as of June  2023. Osculating heliocentric orbital elements at epoch
 MJD 60,000.0 from the MPC catalog: semimajor axis $a$, eccentricity $e$, inclination $I$,
 longitude of node $\Omega$, and argument of perihelion $\omega$. singleopposition orbits are
 listed at the end of the table. The third column gives the absolute magnitude $H$. The last
 column indicates, whether the asteroid has been detected by CSS during the phase~2
 operations (Y=yes). We note two very small, singleopposition asteroids 2015~KM284 and
 2015~KG287, very likely members of the Rampo family too. However, their orbits, based on
 observations spanning less than a week, are still very uncertain.} \\
 \hline \hline
 \multicolumn{2}{c}{Asteroid} & \rule{0pt}{2ex} $H$ & $a$ & $e$ & $I$ & $\Omega$ & $\omega$ & CSS \\
	& & (mag) & (au) & & (deg) & (deg) & (deg) & \\ [1pt]
 \hline
 \endfirsthead
 \caption{continued.}\\
 \hline\hline
 \multicolumn{2}{c}{Asteroid} & \rule{0pt}{2ex} $H$ & $a$ & $e$ & $I$ & $\Omega$ & $\omega$ & CSS \\
	& & (mag) & (au) & & (deg) & (deg) & (deg) & \\ [1pt]
 \hline
 \endhead
 \hline
 \endfoot
 \rule{0pt}{3ex}
  10321 & Rampo      & 14.37 & 2.3285978 & 0.0952815 & 6.06091 & 53.88221 & 278.53547 & Y \\    
 294272 & 2007 UM101 & 17.55 & 2.3294915 & 0.0944526 & 6.05304 & 53.16108 & 280.07932 & Y \\    
 451686 & 2013 BR67  & 17.75 & 2.3278183 & 0.0943492 & 6.09428 & 61.69559 & 266.60300 & Y \\    
 546329 & 2010 VO19  & 18.64 & 2.3284414 & 0.0929491 & 6.09152 & 62.49298 & 265.50211 & Y \\    
 562123 & 2015 XH207 & 18.17 & 2.3273763 & 0.0941514 & 6.09700 & 62.36822 & 265.60187 & Y \\    
 601678 & 2013 JF69  & 18.45 & 2.3287200 & 0.0936834 & 6.08492 & 60.24854 & 268.75113 & Y \\    
        & 2005 VO22  & 18.60 & 2.3284379 & 0.0942207 & 6.08237 & 60.44807 & 269.16712 & Y \\    
        & 2006 UA169 & 18.30 & 2.3287767 & 0.0936955 & 6.07318 & 58.37005 & 272.18424 & Y \\    
        & 2007 XP67  & 18.38 & 2.3290438 & 0.0937905 & 6.07805 & 58.76112 & 271.03067 & Y \\    
        & 2008 GZ170 & 18.33 & 2.3278729 & 0.0934523 & 6.08337 & 60.94947 & 268.37592 & Y \\    
        & 2008 SW341 & 18.33 & 2.3299539 & 0.0957301 & 6.04252 & 51.57990 & 282.43371 & Y \\    
        & 2009 HD95  & 18.15 & 2.3289378 & 0.0934420 & 6.08220 & 60.06838 & 269.21119 & Y \\    
        & 2009 SR371 & 18.70 & 2.3287466 & 0.0939626 & 6.06727 & 56.76277 & 274.96649 & Y \\    
        & 2009 WB276 & 18.46 & 2.3282859 & 0.0941812 & 6.06667 & 57.01124 & 274.26618 & \\      
        & 2010 VP264 & 18.71 & 2.3283338 & 0.0926865 & 6.10160 & 64.02540 & 263.22675 & \\                
        & 2011 WC22  & 18.63 & 2.3277415 & 0.0937305 & 6.09825 & 62.49161 & 265.21277 & Y \\              
        & 2012 VE126 & 18.70 & 2.3299965 & 0.0951578 & 6.05462 & 53.54667 & 279.68212 & Y \\      
        & 2013 RL101 & 18.10 & 2.3284038 & 0.0931685 & 6.08778 & 61.63153 & 267.08811 & Y \\      
        & 2013 VC30  & 18.53 & 2.3283365 & 0.0936238 & 6.07791 & 59.32627 & 270.66675 & Y \\      
        & 2013 VE51  & 18.78 & 2.3280528 & 0.0931012 & 6.09217 & 62.47444 & 265.73127 & Y \\      
        & 2014 HS9   & 18.38 & 2.3285282 & 0.0950748 & 6.07653 & 58.51435 & 271.58717 & \\      
        & 2014 HN87  & 19.03 & 2.3279516 & 0.0942568 & 6.09716 & 63.26309 & 264.63747 & \\                  
        & 2014 ST44  & 18.97 & 2.3288418 & 0.0947150 & 6.06330 & 55.71611 & 275.58485 & \\                  
        & 2015 BB184 & 18.71 & 2.3285590 & 0.0927376 & 6.09722 & 63.17079 & 264.60035 & \\                  
        & 2015 HT91  & 18.22 & 2.3277235 & 0.0932915 & 6.08888 & 62.06807 & 266.70756 & Y \\                
        & 2015 TA367 & 18.89 & 2.3291271 & 0.0954163 & 6.05779 & 53.26353 & 279.52356 & \\                  
        & 2015 TM372 & 18.57 & 2.3285477 & 0.0949021 & 6.07459 & 57.66486 & 273.16183 & Y \\  
        & 2015 VK190 & 19.02 & 2.3292421 & 0.0954998 & 6.04585 & 51.82977 & 282.18049 & \\    
        & 2016 GJ353 & 19.20 & 2.3296639 & 0.0942243 & 6.06093 & 54.67358 & 277.07351 & \\
        & 2016 PR196 & 19.36 & 2.3298272 & 0.0945697 & 6.03487 & 50.33266 & 284.58940 & Y \\  
        & 2016 TE87  & 18.09 & 2.3281180 & 0.0941459 & 6.07157 & 57.95426 & 272.79711 & Y \\  
        & 2017 UH21  & 18.38 & 2.3289973 & 0.0933377 & 6.08745 & 60.32819 & 268.78683 & Y \\  
        & 2018 NN9   & 18.82 & 2.3281713 & 0.0946714 & 6.08543 & 59.97434 & 269.59068 & \\    
        & 2018 PS68  & 18.39 & 2.3291862 & 0.0955066 & 6.03521 & 49.93007 & 285.17948 & Y \\  
        & 2019 PC41  & 18.75 & 2.3285442 & 0.0939345 & 6.08000 & 59.64944 & 270.69700 & Y \\ 
        & 2020 PJ53  & 18.90 & 2.3297760 & 0.0940307 & 6.05985 & 54.29035 & 278.19354 & \\         
        & 2021 QC81  & 19.05 & 2.3283427 & 0.0938521 & 6.08798 & 60.49374 & 268.42639 & \\         
        & 2022 QE61  & 18.93 & 2.3301748 & 0.0957688 & 6.03862 & 50.74900 & 283.82069 & Y \\         
        & 2022 QU76  & 19.08 & 2.3294832 & 0.0954286 & 6.05628 & 54.51791 & 278.35866 & \\
        & 2022 QY123 & 18.97 & 2.3282912 & 0.0942403 & 6.09698 & 63.00157 & 264.91167 & \\ [6pt]
           \multicolumn{8}{c}{-- Singleopposition members --} \\
           \rule{0pt}{3ex}
	& 2020 MO19  & 18.70 & 2.3285350 & 0.0928385 & 6.09814 & 63.13642 & 264.21127 & \\   
	& 2022 RX76  & 18.99 & 2.3292702 & 0.0952251 & 6.06717 & 56.80863 & 274.65354 & \\ [2pt]
\hline
\end{longtable}

\onecolumn
\begin{longtable}{rlccccccc}
\caption{\label{wasserburg_members}
 Wasserburg family as of June  2023. Osculating heliocentric orbital elements at epoch
 MJD 60,000.0 from the MPC catalog: semimajor axis $a$, eccentricity $e$, inclination $I$,
 longitude of node $\Omega$, and argument of perihelion $\omega$. The third column
 gives the absolute magnitude $H$. The last
 column indicates, whether the asteroid has been detected by CSS during the phase~2
 operations (Y=yes).} \\
 \hline \hline
 \multicolumn{2}{c}{Asteroid} & \rule{0pt}{2ex} $H$ & $a$ & $e$ & $I$ & $\Omega$ & $\omega$ & CSS \\
	& & (mag) & (au) & & (deg) & (deg) & (deg) & \\ [1pt]
 \hline
 \endfirsthead
 \caption{continued.}\\
 \hline\hline
 \multicolumn{2}{c}{Asteroid} & \rule{0pt}{2ex} $H$ & $a$ & $e$ & $I$ & $\Omega$ & $\omega$ & CSS \\
	& & (mag) & (au) & & (deg) & (deg) & (deg) & \\ [1pt]
 \hline
 \endhead
 \hline
 \endfoot
 \rule{0pt}{3ex}
   4765 & Wasserburg & 14.05 & 1.9453591 & 0.0599697 & 23.71330 & 76.50142 & 108.59143 & Y \\ 
 350716 & 2001 XO105 & 18.00 & 1.9459411 & 0.0597860 & 23.70790 & 76.45874 & 108.33248 & Y \\ 
        & 2012 KH56  & 19.22 & 1.9456701 & 0.0604509 & 23.70963 & 76.44694 & 108.32313 & Y \\         
        & 2016 GL253 & 19.18 & 1.9457396 & 0.0598318 & 23.71026 & 76.46677 & 108.53996 & Y \\        
        & 2017 DU131 & 18.90 & 1.9456063 & 0.0604246 & 23.70749 & 76.42791 & 108.29692 & Y \\         
        & 2017 KO46  & 19.27 & 1.9453538 & 0.0604115 & 23.70825 & 76.50900 & 108.09233 & Y \\         
        & 2018 YF16  & 18.94 & 1.9454573 & 0.0602688 & 23.70620 & 76.40472 & 108.19370 & Y \\         
        & 2020 HF21  & 19.01 & 1.9455092 & 0.0604480 & 23.70669 & 76.45580 & 108.27241 & Y \\ [2pt]   
\hline
\end{longtable}

\onecolumn
\begin{longtable}{rlccccccc}
\caption{\label{martes_members}
 Martes family as of June  2023. Osculating heliocentric orbital elements at epoch
 MJD 60,000.0 from the MPC catalog: semimajor axis $a$, eccentricity $e$, inclination $I$,
 longitude of node $\Omega$, and argument of perihelion $\omega$. The third column
 gives the absolute magnitude $H$. The last
 column indicates, whether the asteroid has been detected by CSS during the phase~2
 operations (Y=yes).} \\
 \hline \hline
 \multicolumn{2}{c}{Asteroid} & \rule{0pt}{2ex} $H$ & $a$ & $e$ & $I$ & $\Omega$ & $\omega$ & CSS \\
	& & (mag) & (au) & & (deg) & (deg) & (deg) & \\ [1pt]
 \hline
 \endfirsthead
 \caption{continued.}\\
 \hline\hline
 \multicolumn{2}{c}{Asteroid} & \rule{0pt}{2ex} $H$ & $a$ & $e$ & $I$ & $\Omega$ & $\omega$ & CSS \\
	& & (mag) & (au) & & (deg) & (deg) & (deg) & \\ [1pt]
 \hline
 \endhead
 \hline
 \endfoot
 \rule{0pt}{3ex}
  5026 & Martes     & 14.10 & 2.3785050 & 0.2419535 & 4.28293 & 304.74872 & 17.59761 & Y \\        
       & 2005 WW113 & 17.92 & 2.3766591 & 0.2431729 & 4.29300 & 304.86627 & 17.41694 & Y \\        
       & 2010 TB155 & 17.90 & 2.3771299 & 0.2421036 & 4.28760 & 304.75606 & 17.10483 & Y \\        
       & 2011 RF40  & 19.87 & 2.3771146 & 0.2442609 & 4.29445 & 304.60737 & 17.41062 & \\        
       & 2022 QB59  & 20.10 & 2.3770235 & 0.2441532 & 4.29430 & 304.61079 & 17.40973 & \\        
       & 2022 RM50  & 20.13 & 2.3769466 & 0.2440863 & 4.29433 & 304.61266 & 17.38851 & \\ [2pt]  
\hline
\end{longtable}

\onecolumn
\begin{longtable}{rlccccccc}
\caption{\label{lucascavin_members}
 Lucascavin family as of June 2023. Osculating heliocentric orbital elements at epoch
 MJD 60,000.0 from the MPC catalog: semimajor axis $a$, eccentricity $e$, inclination $I$,
 longitude of node $\Omega$, and argument of perihelion $\omega$. The third column
 gives the absolute magnitude $H$. The last
 column indicates, whether the asteroid has been detected by CSS during the phase~2
 operations (Y=yes).} \\
 \hline \hline
 \multicolumn{2}{c}{Asteroid} & \rule{0pt}{2ex} $H$ & $a$ & $e$ & $I$ & $\Omega$ & $\omega$ & CSS \\
	& & (mag) & (au) & & (deg) & (deg) & (deg) & \\ [1pt]
 \hline
 \endfirsthead
 \caption{continued.}\\
 \hline\hline
 \multicolumn{2}{c}{Asteroid} & \rule{0pt}{2ex} $H$ & $a$ & $e$ & $I$ & $\Omega$ & $\omega$ & CSS \\
	& & (mag) & (au) & & (deg) & (deg) & (deg) & \\ [1pt]
 \hline
 \endhead
 \hline
 \endfoot
 \rule{0pt}{3ex}
  21509 & Lucascavin & 15.06 & 2.2804908 & 0.1126543 &  5.98061 & 70.14627 &  4.71189 & Y \\
 180255 & 2003 VM9   & 17.21 & 2.2806359 & 0.1126433 &  5.98101 & 70.37821 &  4.12878 & Y \\
 209570 & 2004 XL40  & 17.24 & 2.2815589 & 0.1114265 &  5.97968 & 69.95755 &  4.92455 & Y \\ [2pt]
\hline
\end{longtable}

\end{appendix}

\end{document}